\documentclass[5p]{elsarticle}
\usepackage{hyphenat}
\usepackage[T1]{fontenc}
\usepackage[utf8]{inputenc}
\usepackage{amsmath}
\usepackage{amssymb}
\usepackage{bm}
\usepackage{graphicx}
\usepackage{booktabs}
\usepackage{array}
\usepackage{tabularx}
\usepackage{microtype}
\usepackage{hyperref}
\usepackage{siunitx}
\biboptions{sort&compress}

\newcolumntype{P}[1]{>{\raggedright\arraybackslash}p{#1}}
\begin{document}

\begin{frontmatter}

\title{Towards Cavity-Based X-ray Free-Electron Lasers: Milestones and Challenges}
\author[addr3]{Kai Li}
\author[addr2]{Patrick Rauer}
\ead{patrick.rauer@desy.de}
\author[addr1]{Nanshun Huang}
\ead{huangnanshun@sari.ac.cn}
\author[addr1]{Haixiao Deng}
\ead{denghx@sari.ac.cn}

\affiliation[addr3]{organization={Department of Chemistry, Princeton University},
            city={Princeton},
            postcode={08544},
            country={USA}}
\affiliation[addr2]{organization={Deutsches Elektronen-Synchrotron DESY},
             city={22603 Hamburg},
             country={Germany}}
\affiliation[addr1]{organization={Shanghai Advanced Research Institute, Chinese Academy of Sciences},
             city={Shanghai},
             postcode={201210},
             country={China}}

\begin{abstract}
Cavity-based X-ray free-electron lasers (CBXFELs) could advance X-ray science by delivering fully coherent radiation with high spectral brightness and enhanced pulse control, addressing the limitations of conventional self-amplified spontaneous emission paradigms. Recent experiments at the European XFEL have demonstrated spectral narrowing and proof-of-concept multi-pass amplification, marking the first experimental signature of the tightly coupled three-body interaction between relativistic electron beams, FEL gain dynamics, and high-finesse Bragg cavities. This review consolidates advances in CBXFEL theory built on this framework, global R$\&$D efforts, and outstanding technical challenges arising from the coupled system. CBXFELs support applications in X-ray spectroscopy, quantum optics, and precision metrology. Future directions prioritize CBXFEL stabilization, X-ray comb generation, and applications in coherent quantum control of nuclear transitions.
\end{abstract}


\end{frontmatter}

\section{Introduction}
\label{sec:introduction}

The free-electron laser (FEL) was originally conceived as a source of stimulated radiation generated by relativistic electrons propagating through a periodic magnetic field \citep{Madey1971}. Unlike conventional lasers, in which amplification relies on bound electronic transitions in atoms, molecules, or solids, FELs use free relativistic electrons as the gain medium. Their resonant wavelength is governed primarily by the electron-beam energy, the undulator period, and the undulator parameter, which makes FELs tunable across a wide spectral range. The advent of high-gain X-ray FELs extended this principle to angstrom wavelengths, where conventional resonator concepts based on efficient normal-incidence mirrors are no longer available and radiation is typically amplified in a single pass. The first angstrom-wavelength lasing at the Linac Coherent Light Source, followed by the operation of other X-ray FEL facilities, established XFELs as femtosecond, high-brightness, transversely coherent sources for probing matter at atomic length scales and ultrafast time scales \citep{Emma2010,Ishikawa2012}; comprehensive reviews of the development and physics of XFELs are given in \citep{McNeil2010,Pellegrini2016}.

The dominant operating mode of early hard-X-ray FELs has been self-amplified spontaneous emission (SASE). In SASE, no external seed at the target wavelength is required: microscopic shot noise in the electron beam is exponentially amplified in a long undulator until saturation \citep{Bonifacio1984,Saldin2000}. This noise-started mechanism gives SASE its simplicity, robustness, and broad applicability, and it has therefore become the foundation of modern XFEL operation. The same mechanism, however, also defines its intrinsic limitations. Because each pulse originates from a different realization of shot noise, SASE radiation exhibits stochastic temporal and spectral substructures, limited longitudinal coherence, relatively broad bandwidth, and significant pulse-to-pulse fluctuations. Thus, although SASE XFELs can deliver highly transversely coherent and extremely bright pulses, their longitudinal coherence and spectral reproducibility remain well below those expected from an ideal transform-limited X-ray laser.

A range of single-pass methods has been developed to improve the coherence, spectral brightness, and temporal control of XFEL pulses. Self-seeding narrows the spectrum by generating a monochromatic seed within a single electron bunch and then amplifying it downstream \citep{Geloni2012Self,Serkez2013Grating,Yang2013Maximizing}. Fresh-slice operation, harmonic lasing, undulator tapering, reverse tapering, and advanced undulator control further extend the accessible parameter space in pulse duration, spectrum, polarization, and peak power \citep{Schneidmiller2016First,Zhang2017Extending,Emma2015High}. These techniques have substantially improved the performance and flexibility of single-pass XFELs. Nevertheless, they do not provide pulse-to-pulse optical feedback. Their performance is therefore still governed by accelerator stability, shot-to-shot electron-beam variations, and single-pass gain dynamics, rather than by the stabilizing action of a resonant optical field.

Cavity-based X-ray free-electron lasers (CBXFELs) seek to introduce such feedback in the hard-X-ray regime by combining FEL gain with Bragg-reflecting crystal cavities and precisely synchronized X-ray recirculation. In an X-ray FEL oscillator (XFELO), a high-$Q$ crystal cavity stores a narrow-band X-ray field and returns it to overlap with successive low-gain electron bunches in an undulator \citep{Kim2008,KimShvydko2009}. In a regenerative-amplifier FEL (RAFEL), a high-gain undulator provides substantial amplification during each pass, while a lower-$Q$ X-ray cavity feeds back only a small fraction of the radiation field \citep{HuangRuth2006,Tang2023}. These concepts occupy different regions of the gain--feedback parameter space. The XFELO emphasizes narrow-band storage, mode selection, and oscillator-like stability, whereas the RAFEL relaxes the cavity-storage requirement by relying on stronger single-pass amplification.

The scientific motivation for CBXFELs arises from the demand for X-ray pulses that combine high spectral brightness, well-defined coherence, and high repetition rate. In this review, we distinguish several related but physically different forms of coherence. Transverse coherence determines the spatial wavefront quality and is essential for coherent imaging and diffraction \citep{Nugent2009Coherent}. Longitudinal coherence is associated with narrow bandwidth and long temporal coherence and is essential for high-resolution spectroscopy and precision measurements. Pulse-to-pulse stability governs reproducibility in pulse energy, spectrum, and arrival time. Phase stability is required for interferometric, X-ray comb, and other phase-sensitive X-ray methods. Many frontier experiments do not merely require more photons; they require photons concentrated into a narrow bandwidth, stable in energy and phase, and delivered in reproducible pulse trains. Nuclear resonant scattering, high-resolution resonant inelastic X-ray scattering, X-ray interferometry, precision spectroscopy, and selected nonlinear X-ray experiments can directly benefit from meV-scale bandwidths and stable pulse sequences \citep{Adams2019,Chubar2015Novel,Gerharz2025Single,Fuchs2015Nonlinear}. Structural biology, coherent imaging, and ultrafast condensed-matter studies can also benefit from increased coherent flux, because it can improve signal per dose, support correlation-based measurements, and enable pump--probe protocols that are difficult to realize with fluctuating single-pass pulses \citep{Chapman2011,Shinohara2020Split}.

The perspective of this review is deliberately integrative. A CBXFEL is not simply a SASE undulator equipped with mirrors, nor is it a conventional laser translated to shorter wavelength. Its gain medium is replenished bunch by bunch, its cavity response is narrow-band and dispersive, and its stored field is sensitive to accelerator jitter, crystal deformation, wakefields, thermal loading, cavity alignment, and user out-coupling \citep{BattermanCole1964,Authier2001,HuangDeng2020Thermal,Liu2024Thermoelastic}. Consequently, the design of a CBXFEL must be treated as a coupled problem involving FEL gain physics, X-ray crystal optics, accelerator stability, diagnostics, feedback control, and application requirements. Throughout this review, we use \emph{CBXFEL} as a generic term encompassing cavity-based X-ray FEL architectures, while distinguishing XFELOs, RAFELs, and oscillator-seeded high-gain amplifiers according to their respective gain mechanisms, optical feedback strategies, and longitudinal mode-selection regimes \citep{Kim2008,HuangRuth2006,DengFeng2013,Adams2019}.

\begin{figure}[hbt!]
    \centering
    \includegraphics[width=\linewidth]{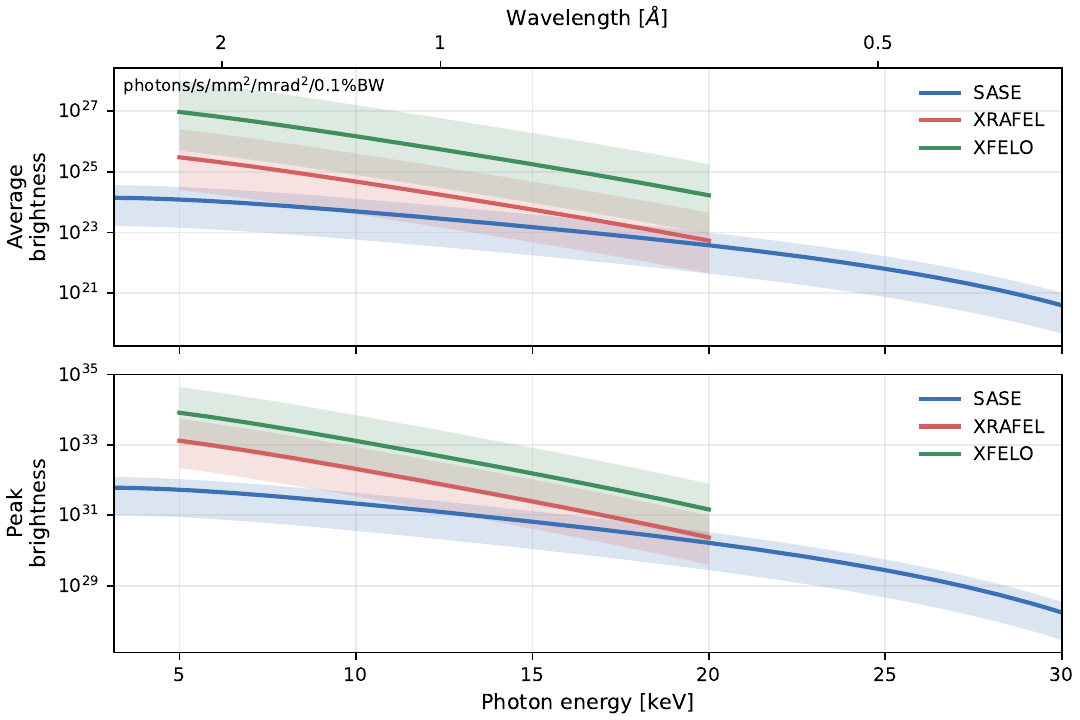}
    \caption{Estimated average and peak spectral-brightness regimes for SASE, RAFEL, and XFELO operation at 8 GeV beam energy, 1 kA peak current, 100 pC bunch charge, and 1 MHz repetition rate. Shaded bands indicate approximate uncertainty ranges arising from assumptions about output coupling, bandwidth, and pulse duration.}
    \label{fig:fel_brightness_regimes}
\end{figure}

The recent first demonstration of lasing with multi-pass gain in a hard-X-ray Bragg cavity marks a milestone for the field, showing that recirculated X-ray pulses can be amplified over repeated passes in an integrated facility-scale experiment \citep{Rauer2026}. This result moves CBXFELs from a largely conceptual and component-demonstration stage toward an experimentally grounded source architecture. It also indicates that several key ingredients, including low-loss crystal cavities, precision timing, X-ray recirculation, and gain-over-loss operation, have reached a level of maturity sufficient for integrated demonstration. At the same time, CBXFELs remain an evolving technology rather than a finished source platform. Critical issues such as thermal loading, crystal deformation, cavity alignment and stabilization, spectral control, out-coupling, wavelength tunability, high-average-power operation, and the preservation of coherence in oscillator-seeded amplification still require further theoretical, experimental, and engineering advances. The brightness regimes summarized in Fig.~\ref{fig:fel_brightness_regimes} illustrate the source-performance gap that motivates cavity-based operation.

Against this background, Section~\ref{sec:theory} establishes the gain physics, cavity metrics, CBXFEL architectures, beam requirements, and simulation methods used throughout this review. Section~\ref{sec:cavity} then examines Bragg reflection, cavity geometry, focusing, thermal load, crystal damage, and diagnostics. Section~\ref{sec:status} assesses CBXFEL activities at major superconducting-linac facilities and in proof-of-principle experiments. Sections~\ref{sec:applications} and \ref{sec:future} consider scientific applications and future directions, respectively. These directions include alignment control, thermal management, pulse shaping, mode locking, out-coupling, and storage-ring implementations. Section~\ref{sec:conclusion} identifies the system-level requirements for user-ready CBXFEL operation.

\section{Theoretical Foundations}
\label{sec:theory}

\subsection{Basic FEL theory}

A free-electron laser relies on resonant energy exchange among a relativistic electron beam, a periodic undulator, and a co-propagating radiation field. In a planar undulator, the transverse electron motion beats with the optical field to form a slowly varying ponderomotive wave. For undulator period $\lambda_u$ and strength parameter $K=eB_0/(mck_u)$, the relevant phase is
\begin{equation}
    \psi=(k+k_u)z-\omega t+\phi .
\end{equation}
Sustained energy exchange requires this phase to remain nearly synchronous with the electron motion. During each undulator period, the radiation therefore slips ahead of the electron by one resonant wavelength. For on-axis radiation at harmonic number $h$, the resonance condition is
\begin{equation}
\lambda_h
=
\frac{\lambda_u}{2h\gamma_r^2}
\left(1+\frac{K^2}{2}\right),
\label{eq:resonance-harmonic}
\end{equation}
where $h=1$ gives the fundamental wavelength. This condition and the ponderomotive phase underpin both low-gain oscillator theory and high-gain FEL instability theory.

Near resonance, the relative energy deviation $\eta=(\gamma-\gamma_r)/\gamma_r$ changes the phase slip, while the optical field changes the electron energy. Averaging over the undulator period yields the standard phase--energy equations
\begin{align}
\frac{d\psi}{dz}
&=
2k_u\eta,
&
\frac{d\eta}{dz}
&=
-a_s\sin\psi.
\label{eq:pendulum}
\end{align}

Rather than independently interacting with the undulator and radiation fields, electrons are modulated by the combined ponderomotive wave with $a_s$ the field-dependent coupling strength. Trapped electrons execute synchrotron-like oscillations within a ponderomotive bucket, transferring energy according to their phase. Electrons at decelerating phases transfer kinetic energy to the radiation, whereas those at accelerating phases absorb energy. Net amplification requires a phase--energy correlation for which more electrons lose energy than gain it, after averaging over the beam.

\subsection{Low-gain solution and Madey-type gain curve}

In the low-gain regime, the radiation field changes little during one undulator pass and can be treated as an externally specified field. Define the phase mismatch as
\begin{equation}
\delta=\frac{d\psi}{dz},
\qquad
\nu=\delta L_u ,
\end{equation}
where $L_u=N_u\lambda_u$ is the undulator length. For an initially uniform phase distribution, the first-order energy modulation averages to zero. However, it alters the subsequent phase evolution and thereby produces a second-order mean energy exchange. The resulting small-signal power gain has the universal detuning dependence
\begin{equation}
G(\nu)=G_0\,g(\nu),
\qquad
g(\nu)
=
-\frac{d}{d\nu}
\left[
\frac{\sin(\nu/2)}{\nu/2}
\right]^2 .
\label{eq:low-gain-curve}
\end{equation}
Depending on the detuning convention for $\nu$, the sign of the standard Madey-type FEL gain curve in Equation~\eqref{eq:low-gain-curve} changes. The curve makes the gain proportional to the derivative of the spontaneous undulator-radiation spectrum and provides the theoretical basis for low-gain FEL oscillators.

In an oscillator, the single-pass small-signal gain must exceed the total round-trip cavity loss. With $R_{\rm cav}=1-L_{\rm rt}$ denoting the round-trip power-survival factor, the pass-to-pass power map is
\begin{equation}
P_{n+1}
=
R_{\rm cav}\,[1+G(P_n)]P_n
+
P_{\rm sp},
\label{eq:pass-to-pass-map}
\end{equation}
where $P_{\rm sp}$ represents start-up from spontaneous radiation. For $L_{\rm rt}\ll 1$, the approximate threshold is
\begin{equation}
G_0>L_{\rm rt}.
\end{equation}
The oscillator saturates when electron trapping and induced energy spread reduce the large-signal gain to the cavity loss.

\subsection{High-gain solution}

The low-gain map becomes insufficient when the optical field grows appreciably within one undulator pass. In the high-gain regime, the radiation field, electron phase space, and microbunching evolve self-consistently. Under the slowly varying envelope approximation,
\begin{align}
\left(
\frac{\partial}{\partial z}
+
\frac{1}{c}\frac{\partial}{\partial t}
\right)\mathcal{E}
&\propto
n_b K[JJ]\,b,
\label{eq:field-evolution}
\\
b&=\left\langle e^{-i\psi}\right\rangle ,
\label{eq:bunching-factor}
\end{align}
where the bunching factor $b$ measures coherent density modulation at the radiation wavelength. A random beam has $|b|\sim N_\lambda^{-1/2}$, whereas a microbunched beam can approach $|b|\sim 1$.

The FEL Pierce parameter $\rho$ provides the natural high-gain normalization. A useful approximation is
\begin{equation}
\rho
\simeq
\frac{1}{2\gamma_r}
\left(\frac{I}{I_A}\right)^{1/3}
\left(
\frac{K[JJ]\lambda_u}{2\pi\sigma_\perp}
\right)^{2/3},
\label{eq:pierce-parameter}
\end{equation}
where $I_A\simeq 17\,\mathrm{kA}$ is the Alfv\'en current and $\sigma_\perp$ is the rms transverse beam size. Geometry and mode-overlap factors alter the numerical prefactor but not the physical role of $\rho$. It sets the gain bandwidth, tolerable energy spread, saturation efficiency, cooperation length $L_c=\lambda_r/(4\pi\rho)$, and natural high-gain scaling length.

With Eq.~\ref{eq:field-evolution} the linearized self-consistent FEL equations have an exponentially growing eigenmode. The radiation power then has the standard high-gain form
\begin{align}
P(z)
&\simeq
\frac{P(0)}{9}
\exp\left(\frac{z}{L_{G,1D}}\right),
\label{eq:power-growth}
\\
L_{G,1D}
&=
\frac{L_{\rm sc}}{\sqrt{3}}
=
\frac{\lambda_u}{4\pi\sqrt{3}\rho}.
\label{eq:gain-length-1d}
\end{align}

Three-dimensional effects lengthen the gain length relative to Equation~\eqref{eq:gain-length-1d}, including diffraction, emittance, betatron motion, energy spread, wakefields, and finite transverse overlap. In the nonlinear regime, exponential growth stops when induced energy spread and particle trapping become comparable to the FEL bucket scale. The usual high-gain estimates are
\begin{equation}
P_{\rm sat}
\sim
\rho P_{\rm beam},
\qquad
\frac{\Delta\omega}{\omega}
\sim
\rho,
\qquad
L_{\rm sat}
\sim
20\,L_G .
\label{eq:saturation-scalings}
\end{equation}
The same exponential law is commonly used in RAFEL pass-to-pass models. The saturation-bandwidth and efficiency scalings are standard high-gain X-ray FEL results.

Because the FEL gain bandwidth is of order $\rho$, the high-gain scalings require the uncorrelated slice energy spread to satisfy
\begin{equation}
\frac{\sigma_\gamma}{\gamma_r}
\lesssim
\rho .
\label{eq:energy-spread-requirement}
\end{equation}
A correlated energy chirp must also remain small across the cooperation length and the portion of the bunch contributing to gain:
\begin{equation}
\left|
\frac{1}{\gamma_r}\frac{d\gamma}{ds}
\right|L_c
\lesssim
\rho .
\label{eq:chirp-requirement}
\end{equation}
The transverse-emittance requirement follows from the angular term in the resonance condition. A commonly used diffraction-limited estimate is
\begin{equation}
\epsilon_n
\lesssim
\frac{\gamma_r\lambda_r}{4\pi}.
\label{eq:emittance-requirement}
\end{equation}
For hard-X-ray FELs, this condition places the normalized emittance in the sub-micrometer regime. The beam size must also match the optical mode because reducing $\sigma_\perp$ increases $\rho$, whereas excessive focusing increases angular spread and emittance-induced detuning. Low-gain XFELO designs usually choose the electron beta function and X-ray Rayleigh range to maximize electron--photon mode overlap.

Peak-current requirements depend strongly on the operating regime because high-gain single-pass FELs and RAFELs require large $\rho$. They therefore usually favor high peak currents and short bunches. By contrast, an XFELO can operate with lower single-pass gain because its cavity stores and recirculates the field over many passes. Representative XFELO studies use peak currents of $10\text{--}200\,\mathrm{A}$ and normalized emittances of approximately $0.1\text{--}0.3\,\mu\mathrm{m}$. Other typical parameters include relative energy spreads of a few $10^{-4}$ or less, picosecond bunch durations, megahertz repetition rates, and order-unity single-pass gain \citep{Kim2008,Lindberg2011XFELOPerformance}.

\subsection{Transition to cavity-based X-ray FELs}

The preceding theory separates cavity FELs into two limiting regimes. In an XFELO, the undulator provides low single-pass gain, and the stored field grows only when this gain compensates for round-trip cavity loss. In a RAFEL, the undulator acts as a high-gain amplifier. The cavity returns only a small fraction of the output pulse as a coherent seed for the next electron bunch. Both concepts use optical feedback but balance gain, loss, and spectral selection differently.


An XFELO extends the conventional FEL oscillator concept to hard-X-ray wavelengths. It replaces normal-incidence metallic mirrors with high-reflectivity Bragg crystals, typically diamond. The electron beam and undulator provide low-gain FEL amplification, while the crystal cavity provides feedback, spectral filtering, and transverse-mode selection. Equation~\eqref{eq:pass-to-pass-map} captures how the stored X-ray pulse starts from spontaneous radiation and grows over many round trips. It saturates when the large-signal gain falls to the cavity loss. A useful threshold estimate is
\begin{equation}
G_0>L_{\rm rt}.
\label{eq:xfelo-threshold}
\end{equation}
Kim, Shvyd'ko, and Reiche proposed combining ultralow-emittance, multi-GeV electron beams with a low-loss crystal cavity \citep{Kim2008}. Their XFELO concept generates angstrom-wavelength pulses with transverse and temporal coherence, meV-scale bandwidth, and picosecond duration. Modern designs retain a core architecture comprising a low-emittance electron source, undulators and chicanes, near-unity-reflectivity Bragg crystals, output coupling, and low-aberration focusing optics. Because $G_0$ is modest, XFELO performance is sensitive to cavity loss, crystal reflectivity, mode matching, electron energy spread, and timing overlap. Strong cavity-mode selection and ultranarrow bandwidth make XFELOs attractive for nuclear resonant scattering, inelastic X-ray scattering, high-resolution spectroscopy, and related narrow-band applications.

\begin{figure}[hbt!]
    \centering
    \includegraphics[width=\linewidth]{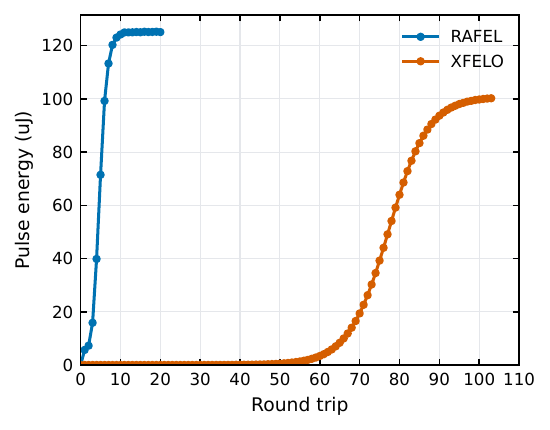}
    \caption{Cavity pulse-energy evolution during the first 100 round trips of XFELO and RAFEL operation. XFELO builds up gradually before saturating after several tens of round trips, whereas RAFEL reaches quasi-steady energy within the first few passes.}
    \label{fig:roundtrip_pulse_energy}
\end{figure}

A RAFEL operates in the opposite limit, with high single-pass undulator gain and reinjection of only a small fraction of the output field:
\begin{align}
R_{\rm fb}G_{\rm hg}
&>1,
\qquad
G_{\rm hg}\gg 1
\label{eq:rafel-threshold}
\end{align}
Because the high-gain factor $G_{\rm hg}$ can be exponentially large, the feedback fraction $R_{\rm fb}$ may be only a few percent. This permits strong out-coupling, lower intracavity fluence, relaxed mirror-reflectivity requirements, and a lower risk of optical damage than in a high-$Q$ oscillator. The Los Alamos RAFEL demonstration used a low-$Q$ optical cavity that reinjected less than 10\% of the optical power \citep{Nguyen1999FirstRAFEL}. It fed this power into a high-gain undulator and reached saturation within a few passes. Figure~\ref{fig:roundtrip_pulse_energy} contrasts the expected pass-to-pass energy buildup in XFELO and RAFEL operation.

\begin{figure}[hbt!]
    \centering
    \includegraphics[width=\linewidth]{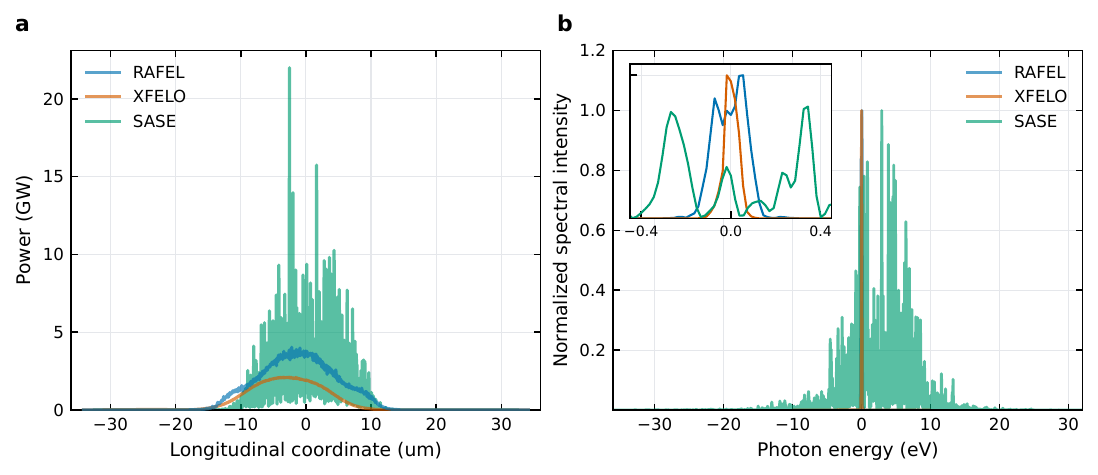}
    \caption{Temporal power and spectral output for 9.8 keV FEL modes. (a) Final longitudinal power profiles for XFELO, RAFEL, and SASE operation. XFELO and RAFEL traces show their steady states, whereas SASE shows the single-pass run. (b) Corresponding normalized spectra. The oscillator modes concentrate output near the design energy, whereas SASE has a much broader spectral envelope.}
    \label{fig:mode_comparison}
\end{figure}

For CBXFELs, narrow-bandwidth Bragg crystals can serve as both feedback optics and spectral filters. The returned field seeds the next bunch. XFELOs are therefore many-pass, low-gain oscillators with fractional single-pass gain comparable to the round-trip loss $L_{\rm rt}<1$. The Bragg cavity primarily shapes XFELO output, making it sensitive to cavity loss, detuning, and wavefront errors. However, the XFELO output can remain stable against shot-to-shot electron-beam fluctuations. RAFELs are few-pass regenerative amplifiers with $G_{\rm HG}$ much greater than unity and $R_{\rm fb}\ll 1$. RAFEL output tolerates modest cavity loss more readily but is more sensitive to beam phase-space jitter, saturation dynamics, and nonlinear wavefront evolution. Figure~\ref{fig:mode_comparison} compares the resulting temporal-power and spectral characteristics of XFELO, RAFEL, and SASE operation.

While the preceding discussion delineates XFELO and RAFEL as distinct operational regimes based on their gain-feedback balance, the boundary between them is not absolute in practice. In a high-repetition-rate XFEL facility equipped with a high-brightness electron beam, the operational mode can be dynamically tuned across a continuum. By adjusting the number of active undulator segments, thereby modulating the single-pass gain, and simultaneously optimizing the coupling efficiency of the X-ray cavity mirrors (e.g., via transmissive diamond beam splitters or angular detuning), a single facility can transition continuously between oscillator-like and regenerative-amplifier-like behavior. This operational flexibility allows the source to identify the optimal working point tailored to specific user demands: prioritizing the high spectral purity and longitudinal coherence characteristic of the XFELO regime for high-resolution spectroscopy or favoring the higher peak power and extraction efficiency of the RAFEL regime for nonlinear X-ray studies. Consequently, future CBXFELs may not be strictly classified as one or the other, but rather as configurable platforms capable of covering the gain-feedback parameter space in real time.

\subsection{Numerical Simulation Tools}
Modeling CBXFELs requires three tightly coupled components: the FEL interaction, X-ray wavefront propagation, and interaction between the radiation and cavity crystals. Unlike SASE simulations, this cycle must be repeated between ten and several hundred times before a CBXFEL reaches saturation. One modeling approach uses approximations to describe multi-pass evolution analytically or semi-analytically. Recent work has advanced this fast approach, which permits broad parameter scans~\cite{Qi2022Misalignment,Tiwari2022Misalignment,Robles2023FastRAFEL}. Depending on the model, assumptions may include low-gain theory, specific electron-bunch shapes, Gaussian radiation profiles, or omission of the FEL radiation source. Such simplifying assumptions are not always applicable under realistic CBXFEL operating conditions.

The second approach uses start-to-end simulations for a more complete treatment. These simulations usually distribute the three coupled components across specialized programs. The X-ray--crystal interaction includes optical reflectivity and the thermal response to absorbed radiation. Crystal reflectivity, discussed in Section~\ref{sec:crystalOpt}, can usually be incorporated directly into the wavefront-propagation calculation.

CBXFEL studies generally model the FEL interaction with the same programs used for other FEL schemes. Common grid-based solvers include the three-dimensional \emph{Genesis-1.3}~\cite{Reiche1999} and the two-dimensional \emph{Ginger}~\cite{Fawley2002Ginger}, while \emph{Medusa} uses Gaussian-mode superpositions~\cite{Freund2016Three}. The output field then enters a wavefront-propagation code such as \emph{OPC}~\cite{Karssenberg2006OPC,Freund2019}, \emph{SRW}~\cite{chubar1998SRW}, or \emph{pXCP}~\cite{Rauer2022PhD}. These calculations must carefully treat dynamical diffraction of the X-ray field by the crystals. Diffraction monochromatizes the incident field and may clip its angular spectrum because of the finite angular acceptance. All three propagation codes include dynamical diffraction, while \emph{pXCP} additionally treats three-dimensional radiation profiles and reflection from imperfect or thermally strained crystals. Other dynamical-diffraction codes include \emph{BRIGHT}~\cite{Huang2019Bright} and the FFT-BPM approach of Krzywinski and Halavanau~\cite{Krzywinski2022BPM}. Both can treat three-dimensional effects, while FFT-BPM can also represent complex crystal deformations.

The crystal thermal response limits the attainable saturation power (see Section~\ref{ssec:thermal}). Thermalization is delayed relative to absorption, allowing separation of the thermal response from dynamical diffraction because a pulse does not affect its own reflection. Dedicated codes model this response using either quasi-analytic approximations~\cite{Zemella2012,HuangDeng2020Thermal} or finite-element analysis~\cite{Rauer2023CBXFELDem,Bahns2024,Zhang2025ThermalDeform}. The resulting deformation can then be included in dynamical-diffraction calculations, for example through FFT-BPM or \emph{pXCP}.

Because this coupled workflow is computationally intensive, CBXFEL studies typically begin with a simplified, fast model. This model optimizes the baseline X-ray optical layout and undulator line before detailed start-to-end simulations finalize the parameter set.

\section{X-ray Optical Cavity}
\label{sec:cavity}
The X-ray optical cavity is the central component of a CBXFEL. Its properties enable the characteristic radiation performance and, for an XFELO, largely define it. Cavity design is therefore a primary consideration when planning a CBXFEL source.
\subsection{Crystal optics}
\label{sec:crystalOpt}
Refraction-based X-ray reflection is restricted to grazing angles on the single-\si{\milli\radian} scale. Hard-X-ray cavities instead use \emph{dynamical diffraction} from single-crystal mirrors at incidence angles extending to near normal incidence. These mirrors both confine and spectrally filter the radiation. This intrinsic filtering distinguishes X-ray cavities from optical resonators and provides the spectral purity of CBXFEL sources. We first summarize the reflection physics and then examine the resulting choice of crystal material.

A crystal irradiated by X-rays of wavelength $\lambda$ reflects strongly when constructive interference matches the scattering vector to a reciprocal lattice vector. For planes indexed by $H = (h,k,l)$, this vector is $\bm{G}_H=h\bm{b}_x+k\bm{b}_y+l\bm{b}_z$, where $b_{x;y;z}$ are reciprocal \emph{Bravais-lattice} basis vectors~\cite{Laue1913}. The same condition is expressed by Bragg's law~\cite{Bragg1913}, %
\begin{equation}
  2 d_H \sin\Theta_B = n\lambda, \qquad n \in \mathbb{Z},
  \label{eq:bragg}
\end{equation}
where $\Theta_B$ is the glancing angle relative to the reflecting planes. For a cubic crystal with lattice constant $a_0$, $d_H=a_0/\sqrt{h^2+k^2+l^2}$. Higher-order diffraction is represented by the corresponding higher-index reciprocal-lattice vector. For an infinite perfect crystal, kinematic theory gives delta-like reflection and cannot predict the high-reflectivity plateau, extinction or phase response. These properties arise from repeated coherent scattering and internal refraction.

Dynamical diffraction accounts for these effects~\cite{Shvydko2004Book}. The internal field comprises Bloch waves coupled through Fourier components $\chi_H$ of the electric susceptibility. In the two-beam approximation, one set of planes satisfies the reflection condition, producing a finite region of near-total reflection and a refractive shift of its centre. For a thick, weakly absorbing, centrosymmetric crystal, the full relative spectral \emph{Darwin width} is approximately
\begin{equation}
  \epsilon_H = \frac{\Delta E}{E_c}
  = \dfrac{4\, r_e\, d_H^2}{\sqrt{|b|}\,\pi V}\, |P F_H|,
  \label{eq:darwinE}.
\end{equation}
Here, $r_e$ is the classical electron radius, $V$ is the unit-cell volume and $P$ is the polarisation factor. The asymmetry factor $b$ describes the orientation of the planes relative to the surface, with $b=-1$ for symmetric Bragg reflection. The structure factor $F_H$ sums the scattering contributions from the unit cell and suppresses forbidden reflections. Asymmetric reflection can tailor the beam size and acceptance, but it also introduces angular dispersion and complicates a multi-element cavity.

\begin{figure}[hbt!]
    \centering
    \includegraphics[width=\linewidth]{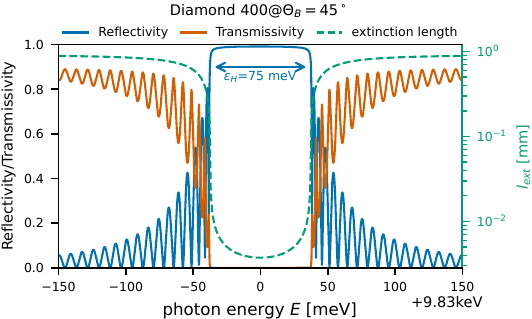}
    \caption{Spectral reflectivity (blue) and transmissivity (orange) of the symmetric diamond 400 reflection at \(\Theta_B=\SI{45}{\degree}\) and \(t_c=\SI{100}{\micro\meter}\). The green curve gives the extinction depth on the logarithmic right-hand axis, which falls to the micrometre scale within the Bragg-reflection band.}
    \label{fig:ReflTrans}
\end{figure}

For diamond and silicon, allowed reflections have either all odd indices or all even indices with $h+k+l$ divisible by four. The Darwin width scales with $|F_H|d_H^2/V$, so higher-order reflections are narrower and have larger extinction depths. Typical crystal thicknesses near \SI{100}{\micro\meter} exceed the in-band extinction depth, making peak reflectivity only weakly dependent on further thickness increases. Figure~\ref{fig:ReflTrans} illustrates these coupled spectral and depth scales for the diamond 400 reflection.
 
For a cavity operated near backscattering, $\Theta_B \approx 90^\circ$, the
angular \emph{Darwin width} broadens via~\cite{Shvydko2004Book}%
\begin{equation}
  \Delta\Theta_H = \begin{cases}
    \epsilon_H \tan\Theta_B & \Theta_B < \frac{\pi}{2},\\
    2\sqrt{\epsilon_H} & \Theta_B\rightarrow \frac{\pi}{2}.
  \end{cases}
  \label{eq:darwin-angular}
\end{equation}
whereas the reflection becomes more confined in absolute energy through $\Delta E=\epsilon_H E_c$. Its center follows the modified Bragg condition,
\begin{equation}
    \begin{array}{r l}
        \lambda_c = &\dfrac{2 d_H \sin\Theta_c}{1 + w_H}\,\leftrightarrow\, E_c = \dfrac{\hbar c}{\lambda_c},\\[2.7ex]
         w_H = &\dfrac{b-1}{2b}\dfrac{4\, r_e\, d_H^2}{\pi V}\, \mathrm{Re}\left(F_0\right)
    \end{array}
  \label{eq:modified-bragg}
\end{equation}
The correction $w_H$ accounts for refraction, which kinematic theory neglects. Two additional effects constrain the operating point. Absorption and finite crystal thickness reduce the peak reflectivity below unity and make the reflection curve slightly asymmetric. At exact normal incidence, cubic symmetry permits several reciprocal lattice vectors to meet the reflection condition simultaneously. The two-beam approximation then breaks down into a strongly perturbed multi-beam pattern.

CBXFEL crystal selection must balance reflectivity, thermal performance, radiation hardness, diffraction geometry, and material quality. The leading candidates for CBXFEL cavities are silicon, diamond, and sapphire. Silicon is widely used in X-ray monochromators because large, nearly perfect crystals are readily available and it has good thermal conductivity and low thermal expansion. Diamond offers the highest Bragg reflectivity, with demonstrated values above $99\%$~\cite{Shvydko2011near100,Shvydko2017Diamondoptics}, together with lower absorption, greater radiation hardness than silicon~\cite{Kolodziej2018DiamondLoad}, and high cryogenic thermal conductivity~\cite{Wei1993CThermal,Olson1993CThermal}. However, its cubic structure introduces multi-beam excitation near normal incidence, so diamond cavities must operate several milliradians from \SI{90}{\degree}. Sapphire, with its trigonal structure, enables backscattering without multi-beam excitation~\cite{Yavas2017Sapphire}, favouring two-crystal cavities, although its thermal and radiation properties are inferior to those of diamond. Both sapphire~\cite{Jafari2017SapphireQuality} and diamond~\cite{POLYAKOV2011DiamondQuality} remain difficult to grow with sufficiently large strain-free regions for low-loss cavities. Current CBXFEL programmes therefore prioritise diamond for its combination of low absorption, high reflectivity, and superior thermal performance.

\subsection{X-ray Optical Cavity Layouts}
\begin{figure}[hbt!]
    \centering
    \includegraphics[width=\linewidth]{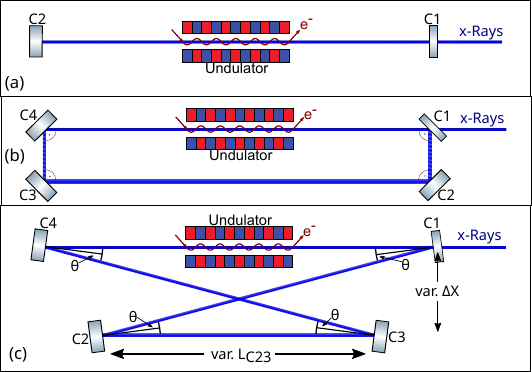}
    \caption{Sketches of the fixed-wavelength (a) two-crystal and (b) rectangular cavities, together with the tunable (c) bow-tie cavity, based on the geometries discussed in Refs.~\cite{Rauer2023CBXFELDem}.}
    \label{fig:CavitySketch}
\end{figure}
A cavity must close the X-ray path while restoring transverse and longitudinal overlap with each successive electron bunch. For one interaction per bunch, the round-trip length is $L=c/f_b$. Megahertz-class bunch rates $f_b$ therefore require paths from several tens to several hundreds of metres, far longer than conventional optical cavities~\cite{Rauer2023CBXFELDem,Rauer2026}. Bragg's law couples photon energy to the glancing angle at each crystal and therefore restricts the available closed paths. The simplest feasible layouts are two-crystal cavities near $\Theta_B\sim$\SI{90}{\degree} and four-crystal cavities near $\Theta_B\sim$\SI{45}{\degree}. These geometries were used in the European XFEL~\cite{Rauer2023CBXFELDem,Rauer2026} and LCLS~\cite{Marcus2019LCLSCBXFELDemo,Margraf2023LowLoss} proof-of-concept experiments, respectively (Fig.~\ref{fig:CavitySketch}).

Near-backscattering two-crystal layouts combine narrow bandwidth, milliradian angular acceptance and a compact return path. Cubic crystals nevertheless require off-axis steering to avoid multi-beam diffraction. The European XFEL cavity operated \SI{4.4}{\milli\radian} from backscattering and used two grazing-incidence Kirkpatrick--Baez mirrors at each crystal. Each crystal--mirror assembly formed a retroreflector that provided passive angular stability~\cite{Rauer2023CBXFELDem,Rauer2026}. This passive stability was central to the successful demonstration at European XFEL discussed in Section~\ref{sec:status}. Four-crystal layouts accept cubic crystals directly and leave more room for diagnostics, but require a larger footprint. In the LCLS implementation, the chosen reflections imposed angular acceptance below \SI{10}{\micro\radian}~\cite{Liu2024Diagnostics}. Their reflection planes must also be oriented consistently with the incident polarisation. The two layouts thus exchange compactness and passive stability against diagnostic access and angular tolerance.

Energy tunability creates a separate design trade-off. FEL photon energy can vary widely through the undulator parameter or electron energy, but fixed-angle cavities remain locked to one central wavelength and its integer harmonics. A grazing-incidence element can, in principle, scan the angle of a two-crystal cavity. A quasi-rectangular four-crystal cavity recently achieved a \SI{10}{\electronvolt} range by combining crystal species with grazing-incidence mirrors~\cite{Koehlenbeck2024Dynamic}. The bow-tie four-crystal geometry offers a wider tuning principle (Fig.~\ref{fig:CavitySketch}(c))~\cite{KimShvydko2009}. It varies crystal angle and transverse displacement together while maintaining a closed cavity of constant length $L$. The required displacements are~\cite{KimShvydko2009}
\begin{align}
    \Delta X = &\frac{L}{2}\tan\left(\Theta_B\right),\\
    L_{C41} + L_{C23} = &\frac{L}{2}\left(1-\tan\left(\Theta_B\right)^2\right).
\end{align}
For $L\sim$\SI{100}{\meter}, tuning from \(\Theta_B\)=\SI{5}{\degree} to \SI{10}{\degree} corresponds to a relative energy change of $\delta E$=\SI{50}{\percent}. The transverse offset then grows from $\Delta X$=\SI{8.75}{\meter} to \SI{17.6}{\meter}. Maintaining sufficient motion accuracy across this range is demanding, which substantially restricts practical spectral tuning. Alternatively, proposed electron-beam shaping and mode-locking schemes could provide single-eV tuning without steering the X-ray optics~\cite{Zhang2025PulseShaping,Huang2025ModeLock}.

Transverse mode matching introduces a trade-off between gain and heat load. As in optical resonators, focusing determines the cavity eigenmode. A CBXFEL must additionally accommodate the finite electron-beam size and natural FEL divergence. Maximum gain and a clean transverse profile favour a radiation intensity width \(\sigma_r\approx\sigma_e\), where \(\sigma_e\) is the electron-beam size in the undulator. At the crystals, by contrast, a larger radiation footprint reduces peak fluence and the resulting thermal response. Gaussian-beam analysis provides the standard starting point for this optimization. The waist is placed in the undulator and propagated through the cavity. For a two-focus system with focal lengths $f_1$ and $f_2$, the cavity is stable when
\begin{align}
    0\leq\left(1-\dfrac{L}{2f_1}\right)\left(1-\dfrac{L}{2f_2}\right)\leq1
\end{align}
Gaussian-beam propagation then gives the waist size \(w_0=2\sigma_r\) and the waist positions $z_{1;2}$ relative to the mirrors:
\begin{align}
	\begin{array}{c}
	w_0^2 = \dfrac{\lambda}{\pi}\dfrac{\sqrt{L\left(2f_1-L\right)\left(2f_2-L\right)\left(2f_1+2f_2-L\right)}}{\left(2f_1+2f_2-2L\right)}.
	\end{array}
\end{align}
Joint optimization within this stability region and against FEL--cavity mode matching yields a unique solution. In a high-gain RAFEL, however, gain guiding can compensate for mode mismatch, and operation without focusing may remain feasible~\cite{Li2019GainGuided}.

Candidate focusing elements include bent grazing-incidence mirrors~\cite{Rauer2023CBXFELDem} and diamond or beryllium compound refractive lenses~\cite{Huang2023MING,Liu2024Diagnostics}. Mirrors introduce little loss and are achromatic, whereas compound refractive lenses are more compact and easier to handle. Both can distort the wavefront and require high manufacturing precision to preserve X-ray pulse quality.

\subsection{Thermal Loading and Damage}
\label{ssec:thermal}
Every Bragg reflection absorbs a fraction $A = 1 - R - T$ of the incident pulse energy, which is converted predominantly into heat. Although silicon, diamond, and sapphire have low absorption coefficients, CBXFEL repetition rates and circulating powers make this deposition substantial. Theory~\cite{Zemella2012,Yang2018,HuangDeng2020Thermal,Rauer2023CBXFELDem,Bahns2024,Zhang2025ThermalDeform} and recent experiments~\cite{Rauer2026} identify heat loading as a limit to stable operation. Both transient single-pulse heating and the average load accumulated across a bunch train contribute. Thermal expansion shifts the lattice spacing and therefore the reflection centre through the modified Bragg condition~\eqref{eq:modified-bragg}. Simultaneously, a nonuniform temperature profile bends the crystal into a thermal bump. The resulting wavefront distortion defocuses the beam and reduces round-trip gain.

\begin{figure}[hbt!]
    \centering
    \includegraphics[width=\linewidth]{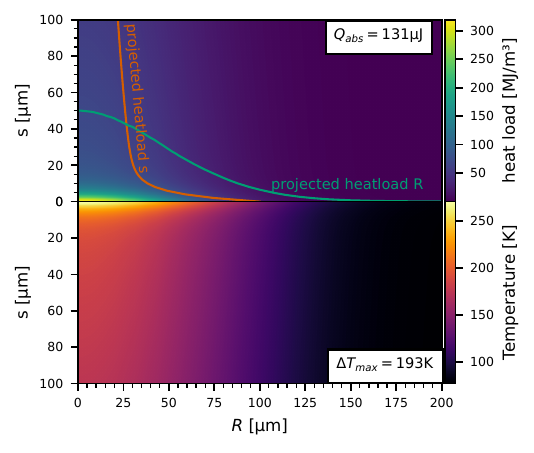}
    \caption{Radially projected heat distribution (top) and subsequent temperature rise from eq.~\eqref{eq:tempjump} (bottom) for a CBXFEL pulse with incident energy $Q_{p}\approx$\SI{1}{\milli\joule}. The simulation uses European XFEL demonstrator parameters and a base crystal temperature of $T_0=$\SI{77}{\kelvin}~\cite{Rauer2026}. The depth profile contains a short-range contribution from radiation within the crystal bandwidth and a long-range contribution from out-of-band radiation. The resulting temperature rise is $\Delta T_{max}\approx$\SI{120}{\kelvin} and must spread through the crystal within one round-trip period.}
    \label{fig:Qabs}
\end{figure}

The radiation width and incidence angle determine the transverse heat distribution, whereas two characteristic lengths set its depth profile. The post-absorption ionization cascade transfers energy to the lattice within picoseconds. Heat diffusion instead occurs over nanoseconds, comparable to the cavity round-trip time. A residual temperature rise can therefore persist until the next pulse. Its instantaneous magnitude follows from the absorbed energy density:
\begin{equation}
  \rho \int_0^{\Delta T(\bm{r})} c_V\left(\delta T'+T_0(\bm{r})\right)\, \mathrm{d}\delta T' = Q_\mathrm{abs}(\bm{r}),
  \label{eq:tempjump}
\end{equation}
where $T_0$ is the pre-absorption temperature, $\rho$ is the mass density, and $c_V$ is the temperature-dependent specific heat. Near saturation, one round trip can deposit more than one hundred microjoules in the diamond crystals. The resulting transient peak temperature rise can exceed one hundred kelvin~\cite{Rauer2023CBXFELDem}. This heat must spread through the crystal within one round trip to preserve cavity performance. Figure~\ref{fig:Qabs} shows an example for a saturated CBXFEL with \SI{1}{\milli\joule} circulating pulse energy, simulated for the European XFEL demonstrator~\cite{Rauer2023CBXFELDem}.
Because specific heat decreases with the base temperature $T_0$, the same thermal load produces a larger temperature rise $\Delta T$ at lower temperatures. A constant thermal-expansion parameter $a_L$ would suggest a correspondingly larger lattice shift through $d_H(T_0+\Delta T)\approx d_H(T_0) + a_L(T_0)\Delta T$. This approximation would appear to disfavour low-temperature operation. The lattice expansion can instead be written as
\begin{align}
        \frac{d_H\left(T_0+\Delta T\right)}{d_H\left(T_0\right)} &=\exp\left(\int_{0}^{\Delta T}\alpha_L(\delta T'+T_0)\mathrm{d}\delta T'\right)-1\notag\\
                                 &=  \exp\left(\int_{0}^{\Delta T}\frac{\gamma_G}{3B}c_V(\delta T'+T_0)\mathrm{d}\delta T'\right)-1  \notag\\ 
                                 & \approx  \exp\left(\frac{\gamma_G}{3B}Q_\mathrm{abs}\right)-1,
\end{align}
The final line uses eq.~\eqref{eq:tempjump} and assumes a material-dependent but temperature-independent mean \emph{Gr\"uneisen} parameter $\gamma_G$ and bulk modulus $B$. Under these approximations, absorption-induced thermal expansion is largely independent of the initial crystal temperature.

Between pulses, thermal diffusion redistributes the deposited heat. This process can be approximated by the transient heat equation
\begin{equation}
  \rho\, c_V(T)\, \frac{\partial T(\bm{r},T)}{\partial t}
  = \nabla\!\cdot\!\left(\bm{\kappa}(T)\, \nabla T(\bm{r},t)\right) + \dot{Q}_\mathrm{gen}(\bm{r},t),
  \label{eq:heateqn}
\end{equation}
where $\bm{\kappa}$ is the temperature-dependent thermal-conductivity tensor and $\dot{Q}_\mathrm{gen}$ is the volumetric heat source from the absorbed pulse. The solution determines whether the interaction volume recovers before the next pulse. Thermal conductivity is therefore the decisive material parameter for pulse-to-pulse recovery.

Across a bunch train, heat accumulation becomes nonlinear optical feedback. In a two-crystal cavity, the downstream crystal intercepts the amplified spectrum and generally heats more strongly, particularly in a RAFEL. The two reflection bands then separate until their overlap and the circulating field decrease. Lower intensity reduces absorption and permits recovery, producing pulse-energy oscillations~\cite{Rauer2023CBXFELDem,Rauer2026}. Temperature-dependent conductivity can amplify the excursion before loss of spectral overlap collapses the field. Abrupt heating also launches persistent strain waves that perturb both reflectivity and wavefront~\cite{Yang2018,Bahns2024,Liu2024Thermoelastic}.

Cryogenic cooling primarily accelerates this recovery rather than suppressing the instantaneous temperature jump. Reduced phonon--phonon scattering raises the conductivity of silicon and diamond by about an order of magnitude near \SI{100}{\kelvin}~\cite{Wei1993CThermal,Olson1993CThermal,Inyushkin2018RecordKappaSi,Inyushkin2023Kappa}. Faster redistribution suppresses the thermal bump and preserves spectral overlap. Simulations consequently predict roughly tenfold higher attainable pulse energy at \SI{77}{\kelvin} than at room temperature~\cite{Rauer2023CBXFELDem}.

Thermal management therefore extends beyond base temperature. Greater thickness can improve lateral heat spreading and reservoir capacity, but it can also reduce transmitted out-coupling. Continuous-wave operation requires the holder and cooling interface to remove at least the average deposited power. Low thermal boundary conductance makes the contact geometry and mounting stress part of the optical design. Peak dose defines a distinct failure boundary. Models of femtosecond excitation place the nonthermal graphitisation threshold near \SI{0.7}{\electronvolt} per atom~\cite{Medvedev2013Graphitisation}. High-average-load tests probe cumulative heating under much lower peak dose~\cite{Kolodziej2018DiamondLoad}. They therefore establish thermal resilience, not immunity to single-pulse damage.

\subsection{Diagnostics}
The long cavity length $L$ and small electron-bunch dimensions impose tight alignment tolerances. The transverse size $\sigma_e$ and longitudinal extent $\sigma_s$ are typically tens of micrometres. Spatial tolerances therefore reach the single-micrometre scale, while angular tolerances range from tens to hundreds of nanoradians. Higher gain relaxes these requirements, but they remain stringent even for a high-gain RAFEL.

Diagnostics must map the circulating X-rays and electron beam with commensurate precision, ideally in a common reference frame. They also occupy a radiation-exposed electron-beam region and must therefore tolerate the local radiation load. For transverse alignment, scintillating screens such as YAG:Ce or fast YAP:Ce can be combined with sensitive cameras and optics. The $\sim$\si{\mega\hertz} X-ray round-trip rate requires either high-speed cameras or fast gating to resolve individual passes. This requirement is especially demanding during startup, when the seed on the undulator axis is weaker than the intracavity SASE background. Because each screen absorbs part of the circulating field, the signal decays rapidly. Successful alignment therefore requires either a sufficiently strong initial signal or a low background.

Fast photodiodes provide complementary, less spatially resolved information. In rectangular cavities, their signal is sensitive to crystal angular misalignment~\cite{Liu2024Diagnostics}. Alternatively, a photodiode can measure transmission through an aperture that acts as a spatial filter and converts beam displacement into intensity changes~\cite{Rauer2026}. This aperture-based signal was the only fast intracavity diagnostic in the European XFEL experiment. It nevertheless proved both essential and sufficient for precise transverse alignment.

Longitudinal synchronization imposes a different limitation. The cavity length must match the electron-bunch repetition period to within single micrometres. Optical alignment can establish an initial length but is generally not accurate enough for final synchronization. A common-reference diagnostic with the required precision for electron and photon arrival times is not yet available. Longitudinal overlap must therefore be found by scanning cavity length while preserving transverse alignment and searching for seeding onset~\cite{Rauer2026,Liu2024Diagnostics}. The electron orbit and arrival time must remain stable on comparable scales. Slow feedback combined with fast feedforward or adaptive feedback provides this stabilization.

Once seeding begins, diagnostics must cover both onset and saturation at megahertz rates and across a large dynamic range. High-precision, pulse-resolved spectrometers are also required \citep{Kujala2020Hirex,Kauchha2025Spectrograph,Huang2023MING}, ideally both inside and downstream of the cavity.

\section{Status of Development}
\label{sec:status}

The facility requirements for a practical CBXFEL include a high-repetition-rate bunch train, sufficient single-pass gain, a photon beamline capable of hosting a mechanically stable Bragg resonator, nondestructive diagnostics for the circulating field, and machine protection compatible with stored intracavity power. Superconducting-linac facilities such as LCLS-II, the European XFEL, and SHINE are well suited because they can deliver many high-quality bunches at MHz-class spacing. Storage-ring-based proposals remain relevant when transverse-gradient or related compensation schemes can recover the narrow gain bandwidth required for low-gain oscillator operation \citep{Decking2020,Galayda2018LCLSII,Li2023TGU}.

CBXFEL development has entered a new phase following the report of \emph{Lasing of a cavity-based X-ray source} at the European XFEL \citep{Rauer2026}. Earlier work had established the theoretical gain and cavity requirements, demonstrated key Bragg-optics components, and developed facility concepts for XFELO and RAFEL operation. This experiment changed the status of the field by closing a critical feedback loop: a hard-X-ray pulse was recirculated in a Bragg cavity, synchronized to fresh electron bunches, amplified over multiple passes, and observed to ring up only under the appropriate cavity-length condition. It represents the first integrated demonstration that FEL gain and an X-ray resonator can operate as a coupled source.

\begin{table*}[!hbt]
\centering
\footnotesize
\caption{CBXFEL development at major global facilities}
\label{tab:readiness_matrix}
\begin{tabular}{P{0.16\linewidth}P{0.24\linewidth}P{0.28\linewidth}P{0.18\linewidth}}
\toprule
Platform & Relevant design basis & Active work & Future work \\
\midrule
LCLS-II & CW superconducting-linac upgrade, high repetition rate, mature hard-X-ray FEL controls, and a route to LCLS-II-HE hard-X-ray operation & LCLS established hard-X-ray SASE user operation; LCLS-II studies define the high-power superconducting platform; a separate LCLS Bragg-cavity experiment demonstrated low-loss X-ray storage without FEL gain \citep{Emma2010,Galayda2018LCLSII,Margraf2023LowLoss} & Closing the Bragg feedback loop with live FEL gain and developing protected routine user modes \\

European XFEL & MHz bunch trains from a superconducting linac, long SASE1 undulator line, multi-instrument photon transport, and pulse-train diagnostics & Demonstrated cavity-based X-ray lasing at 6.952~keV in a 132.8~m diamond Bragg cavity synchronized to 2.23~MHz bunch spacing \citep{Decking2020,Tschentscher2017EuXFEL,Rauer2019Integration,Rauer2026} & Increasing round-trip retention, heat-load stability, output coupling, and reproducibility for user delivery \\

SHINE & CW superconducting-linac concept, soft-to-hard X-ray coverage, and co-designed XFELO/RAFEL studies & MING proposal, polarization and multimode concepts, and beam-dynamics studies for CW XFEL operation \citep{LiDeng2017SCLF,Huang2023MING,Huang2019Polarization,Yan2019Multi,Zhu2024Flexible} & Planned first lasing in autumn 2026; potential integration of CBXFEL, SASE, and self-seeding user programmes \\
\bottomrule
\end{tabular}
\end{table*}

\subsection{LCLS-II}

\begin{figure}[!hbt]
\centering
\includegraphics[width=0.9\linewidth]{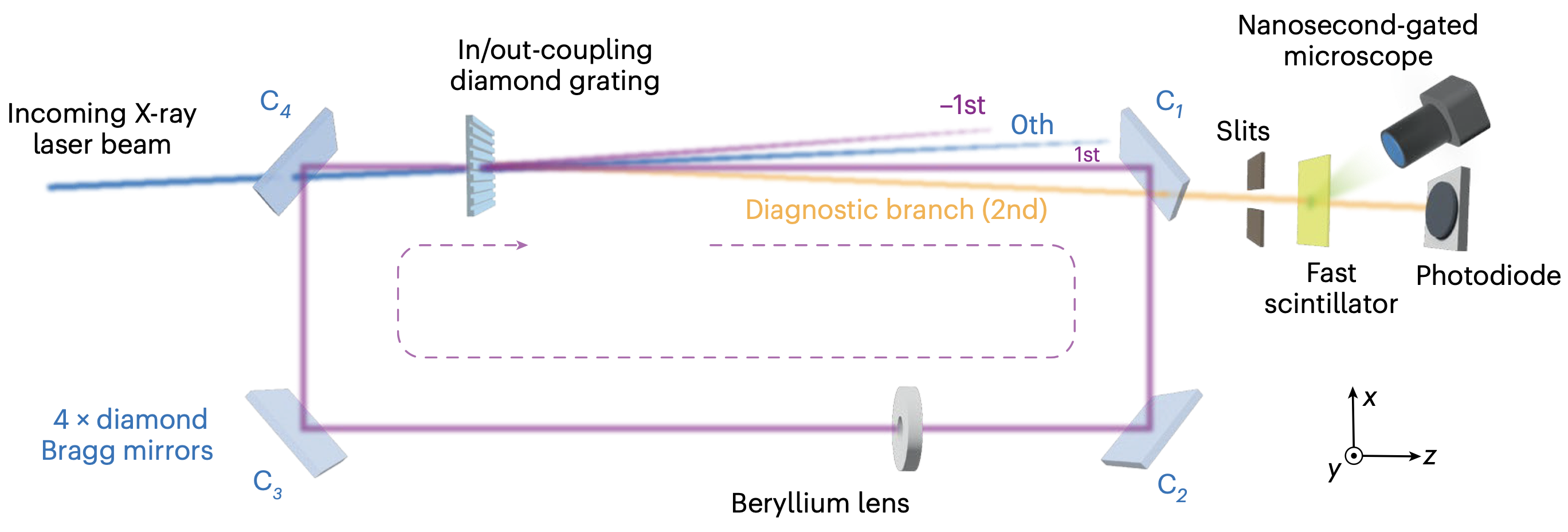}
\includegraphics[width=0.9\linewidth]{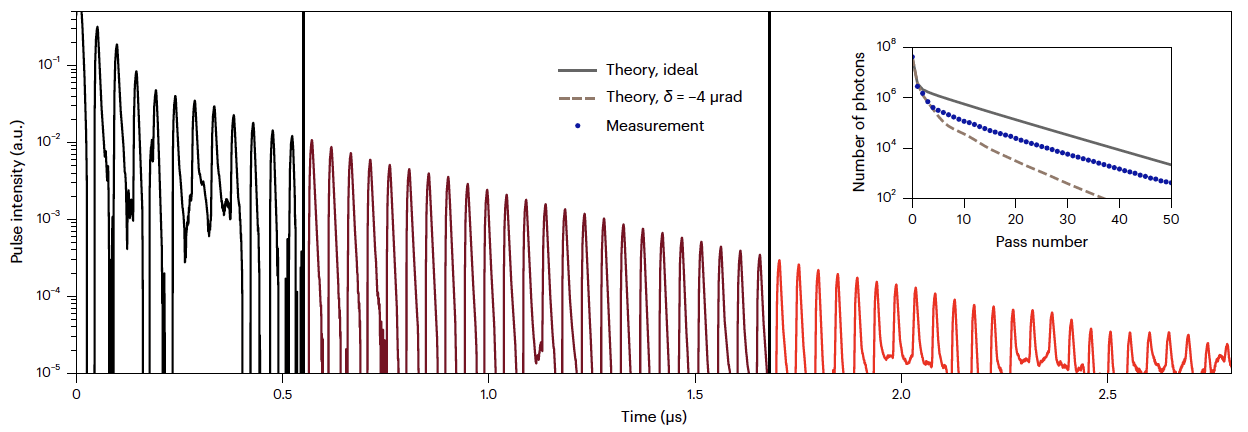}
\caption{LCLS cavity optics and ring-down measurement. Top: Four diamond Bragg mirrors and a transmission grating form the X-ray cavity; the diffracted circulating beam is monitored with a high-speed imager and downstream photodiode. Bottom: Averaged photodiode signals over 59 round trips within 2.8~${\mu}s$, grouped by pass number; the inset shows the estimated intracavity photon number after each round trip. Selected panels adapted from~\cite{Margraf2023LowLoss} under the \href{https://creativecommons.org/licenses/by/4.0/}{Creative Commons Attribution 4.0 International licence}.
\label{fig:LCLC_cold_cavity} 
}
\end{figure}

LCLS provides the historical reference point for hard-X-ray FEL user operation. It demonstrated that angstrom-wavelength SASE pulses could be delivered to a broad scientific program, but also showed the shot-to-shot fluctuations and limited longitudinal coherence characteristic of single-pass startup from noise \citep{Emma2010}. LCLS-II changes the CBXFEL discussion because its superconducting accelerator architecture is designed for high average power and high bunch repetition rate rather than isolated high-charge operation \citep{Galayda2018LCLSII}.

The SLAC group has also provided an important component-level demonstration, seen in Fig.~\ref{fig:LCLC_cold_cavity}. Low-loss storage of 1.2~\AA{} X-ray pulses in a 14~m Bragg cavity at LCLS showed that hard-X-ray pulses can be stored with stable ring-down in a crystal cavity, thereby validating a central optical prerequisite without closing the FEL gain loop \citep{Margraf2023LowLoss}. This distinction is essential. The LCLS cavity experiment tested storage, alignment, loss, and Bragg-optics stability; the later European XFEL lasing experiment tested synchronization of an optical feedback loop to electron bunches with gain. Taken together, these results define a staged experimental logic: first demonstrate that X-rays can survive the cavity, then show that a fresh bunch train can amplify the recirculated pulse, and finally develop controlled output coupling and feedback stabilization for user operation.

For LCLS-II and LCLS-II-HE operation, the practical research questions are now concrete. The cavity length must be locked to the bunch pattern; electron energy and undulator tuning must keep the FEL gain inside the Bragg passband; wakefields and bunch-to-bunch effects must not degrade longitudinal or transverse seed overlap; and the photon line must protect crystals against mis-steered SASE pulses and heat-load transients. A plausible development path is a staged demonstration. At LCLS, ongoing efforts are exploring a twin-bunch configuration to test the coupling between cavity recirculation and FEL gain. In this approach, one bunch can generate or refresh the intracavity X-ray field, while a following bunch interacts with the recirculated pulse, providing a controlled platform to verify multi-pass Bragg-cavity feedback under live FEL amplification.

\subsection{European XFEL}

\begin{figure*}[!hbt]
\centering
\includegraphics[width=0.9\linewidth]{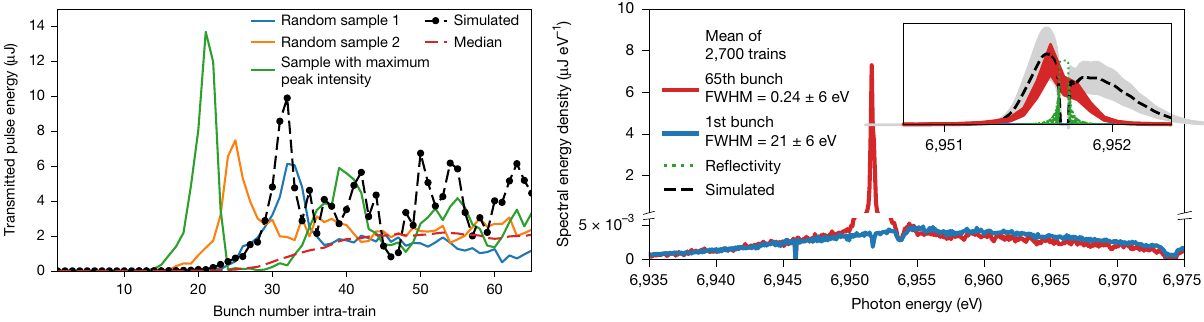}
\caption{Intra-train intensity evolution and spectral narrowing. Left: Integrated spectral intensity versus bunch number for individual trains (colours), with the median over 2,700 trains (red dashed) and a representative simulation (black dashed). Right: Spectra of the first (blue) and 65th (red) bunch downstream of the cavity; inset compares the 65th-bunch spectrum with the ideal downstream-crystal reflectivity (green dotted) and simulation (black dashed). Selected panels combined from Rauer et al.~\cite{Rauer2026} under the \href{https://creativecommons.org/licenses/by/4.0/}{Creative Commons Attribution 4.0 International licence}.\label{fig:Euro_CBXFEL_res}
}
\end{figure*}

The European XFEL is currently a central facility in the experimental CBXFEL record because it was the first to combine a superconducting accelerator, MHz-class intra-train bunch structure, a long hard-X-ray undulator line, and facility-scale photon diagnostics in a single operating environment \citep{Decking2020,Tschentscher2017EuXFEL}. These features made it possible to test a CBXFEL not as an isolated optical-cavity experiment, but as an accelerator-integrated X-ray source.

The Rauer \emph{et al.} demonstration used the SASE1 line at 14~GeV, the last four undulator segments with about 20~m active length, and a 132.8~m round-trip diamond-based Bragg cavity tuned to the 2.23~MHz bunch spacing \citep{Rauer2026}. The lasing photon energy was 6.952~keV, selected by near-backscattering diamond Bragg reflection. Ring-up was observed only when the longitudinal cavity length matched the electron-bunch spacing, providing direct evidence that the recirculated monochromatized pulse seeded successive bunches. The experiment also incorporated in-cavity and downstream diagnostics, including weak grating extraction, spectrometry, imaging, and gas-monitor calibration. 

The result should nevertheless be interpreted with care. It establishes multi-pass gain and spectrally purified microjoule-level output in a real XFEL environment, but it also exposes the engineering limits that must be overcome before routine user operation. The measured round-trip retention remained below ideal expectations based on crystal reflectivity, indicating the importance of mirror aperture, wavefront distortion, crystal strain, and alignment. The pulse-train evolution also revealed heat-load-driven dynamics: as the seeded pulse grew, absorbed power modified the diamond lattice and shifted the Bragg response, creating a feedback loop between gain and thermal detuning \citep{Rauer2026,HuangDeng2020Thermal,Liu2024Thermoelastic}. The European XFEL experiment is therefore both a proof of feasibility and a diagnostic of the next bottleneck: stable high-retention operation with controlled heat load, deliberate output coupling, and reproducible pulse delivery.

\subsection{SHINE}

The SHINE design program is important because it holds the potential to treat cavity-based operation as part of the facility design space rather than only as a later add-on to an existing SASE beamline. The MING proposal at SHINE explicitly frames megahertz cavity-enhanced X-ray generation as a way to exploit an 8~GeV-class CW superconducting linac and generate highly coherent hard-X-ray pulses in the 6--15~keV range \citep{Huang2023MING}. This is precisely the regime in which oscillator bandwidth, Bragg reflectivity, bunch repetition pattern, and user spectral requirements must be optimized together.

SHINE also emphasizes operational flexibility. Polarization control with an XFELO, multi-beam-energy operation for CW XFELs, energy-recovery concepts, and flexible multi-bunch-length operation all address the same facility problem: a high-repetition-rate XFEL must serve many photon modes without sacrificing beam quality \citep{Huang2019Polarization,Yan2019Multi,Wang2020Energy,Zhu2024Flexible}. For CBXFELs, this flexibility is not optional. Different scientific cases may require meV-scale bandwidth, tunable polarization, high pulse energy after amplification, or special bunch timing. SHINE studies therefore provide a design laboratory for co-optimizing accelerator configuration, bunch pattern, undulator staging, crystal choice, cavity round-trip time, and downstream amplifier operation. This inherent versatility aligns with the concept of a continuously tunable gain-feedback system, as introduced previously, positioning SHINE as a prime candidate for demonstrating multi-modal CBXFEL operation.

Compared with the European XFEL, the SHINE CBXFEL mode is less defined by a completed CBXFEL lasing experiment and more by integrated design exploration. Its value is to show how a CBXFEL could be planned from the beginning as one operating mode of a CW hard-X-ray facility. The European XFEL result strengthens the SHINE case because it confirms that the central conceptual step, multi-pass gain in a hard-X-ray cavity, is physically achievable, while leaving open the facility-engineering question of how such operation can be made routine.

\subsection{Experimental Milestones}
As outlined above, there have been two main experimental milestones toward the realization of cavity-based X-ray FELs. These are the passive-cavity, low-loss ring-down experiment at LCLS~\cite{Margraf2023LowLoss} and the recent demonstration of lasing with gain in the demonstrator experiment at the European XFEL~\cite{Rauer2026}.
In addition to demonstrating the basic feasibility of cavity-based X-ray FELs and their fundamental radiation output, including spectral narrowing to $\sim$\SI{200}{\milli\electronvolt} over successive electron bunches, these experiments yielded several further lessons for future work.

Besides confirming the feasibility of high-quality, low-loss diamond-based X-ray optical cavities within the predicted performance range~\cite{Shvydko2017Diamondoptics}, a central lesson from the cold-cavity ring-down experiment was how to transversely align a cavity more than ten meters long using intracavity X-ray diagnostics. An important detail was the impact of the focusing optics on the alignment procedure. These focusing elements introduced a process called betatron oscillation~\cite{Margraf2023LowLoss,Tiwari2022Misalignment}, in reference to the well-known process in accelerator beam optics. In a stable cavity, a misaligned ray executes bounded transverse oscillations about the cavity axis rather than walking out of the aperture, keeping the round-trip loss low even when the elements are imperfectly positioned. The mean trajectory can nonetheless be displaced or tilted relative to the electron beam, so cavity stability does not by itself ensure electron-beam overlap.
The LCLS experiment diagnosed this behavior using fast transverse diagnostics and by moving the CRL lenses into and out of the beam.

The European XFEL lasing demonstration~\cite{Rauer2026} faced similar issues. Because the mirrors were used not only for focusing but also for steering, they could not be removed during alignment. The team therefore relied on a fast photodiode signal combined with a small 200$\times$\SI{200}{\micro\meter\squared} aperture to optimize transverse alignment. Betatron oscillation appeared as small differences between the optimal alignments for the first and higher-order round trips, caused by the betatron phase advance.

The demonstrator experiment at the European XFEL yielded two other important results. First, it demonstrated the importance of a high degree of stability in both the beam parameters (compression, energy, arrival time, and beam trajectory) and the transverse alignment. The latter must be maintained while scanning the cavity length in \SI{1}{\micro\meter} steps over a range of \SI{1}{\milli\meter}. The team achieved this by using the retroreflecting geometry~\cite{Rauer2023CBXFELDem}, which geometrically locked the return beam trajectory while the downstream vacuum chamber was scanned.
Another possibility, also investigated at LCLS, is the use of active stabilization~\cite{Koehlenbeck2024Dynamic}.

Second, the experiment showed clear indications of thermal saturation and pulse-energy beating induced by the strong heat load on the uncooled diamond crystals~\cite{Rauer2026}. Consistent with predictions~\cite{Rauer2023CBXFELDem,Rauer2022PhD}, this observation experimentally highlights the importance of proper cooling and thermal management (see Section~\ref{ssec:thermal}).
\section{Scientific Applications}
The XFELO is expected to deliver X-ray pulses with spectral brightness more than an order of magnitude higher than those of current SASE XFELs, while simultaneously providing orders-of-magnitude improvements in temporal coherence and spectral purity. These properties will directly benefit both linear and nonlinear X-ray spectroscopies, particularly photon-starved and coherence-dependent techniques such as resonant inelastic X-ray scattering (RIXS) and stimulated X-ray Raman spectroscopy (SXRS). The proposed XFELO schemes could further enable coherent multipulse experiments and multidimensional X-ray spectroscopies analogous to those that have advanced optical spectroscopy. Beyond spectroscopy, XFELOs provide a suitable platform for extending the concepts of quantum optics into the X-ray regime, including coherent manipulation of nuclear transitions and the emerging field of nuclear quantum optics. More broadly, the evolution of optical lasers from stochastic light sources to highly coherent frequency-comb systems advanced precision metrology and high-resolution spectroscopy. Likewise, XFELOs offer the prospect of generating phase-coherent X-ray frequency combs. By extending frequency-comb techniques to X-ray wavelengths that are four orders of magnitude shorter than those of optical lasers, XFELOs could advance precision X-ray metrology, high-resolution spectroscopy, and tests of fundamental physics.
\label{sec:applications} 

\subsection{X-ray spectroscopy}
\begin{figure*}[!hbt]
\centering
\includegraphics[width=0.8\linewidth]{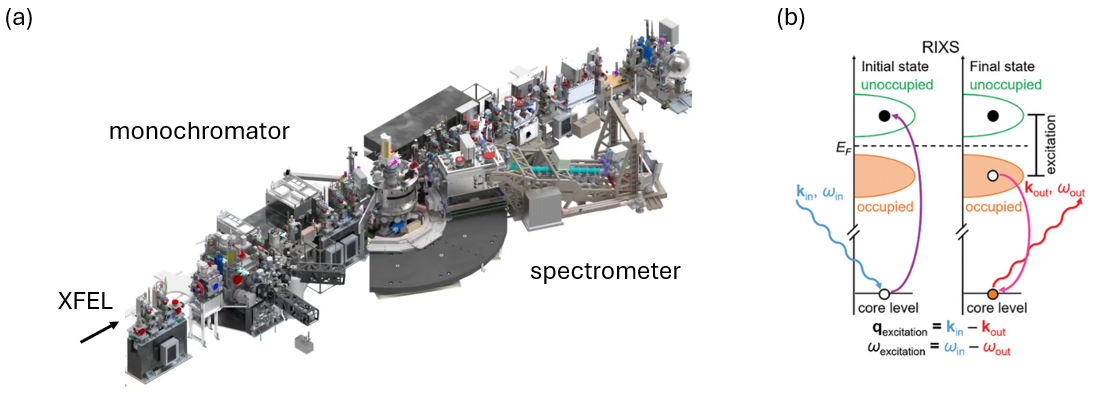}
\caption{\label{fig:rixs} 
(a) Schematic of the chemRIXS setup at LCLS-II~\cite{hoffman2026chemrixs}.
(b) The photon-in/photon-out process of RIXS~\cite{asmara2024emergence}.
}
\end{figure*}
The advent of ultra-bright femtosecond XFEL pulses has created unprecedented opportunities for time-resolved and operando studies of ultrafast dynamics, nonequilibrium processes, and chemical reactions~\cite{loh2020observation,li2024attosecond,bergmann2021using}. For condensed-phase systems of particular relevance to materials science, chemistry, and biology, X-ray photon spectroscopies, including X-ray absorption spectroscopy (XAS), X-ray emission spectroscopy (XES), and resonant inelastic X-ray scattering (RIXS), are among the most powerful tools available due to their elemental specificity, bulk sensitivity, and ability to probe electronic structure. These techniques detect photons emitted, transmitted, or scattered following the interaction of incident X-rays with a sample. In particular, XAS and RIXS measure the outgoing photon signal as a function of incident photon energy and therefore traditionally rely on high-resolution monochromators to produce spectrally narrow incident X-ray beams. As shown in Fig.~\ref{fig:rixs}(a), such monochromators often occupy substantial beamline space, introduce significant complexity, and dramatically reduce the available photon flux. The intrinsically narrow bandwidth and high spectral purity of XFELO radiation could eliminate or substantially relax the need for additional monochromatization, simplifying experimental setups while simultaneously increasing the usable photon flux.

For conventional SASE XFELs and synchrotron sources, monochromatization typically reduces the available flux by one to several orders of magnitude, posing a major challenge for photon-starved spectroscopic techniques. Among these, resonant inelastic X-ray scattering (RIXS) is particularly demanding. As shown in Fig.~\ref{fig:rixs}(b), RIXS is a photon-in/photon-out spectroscopy that combines the strengths of XAS and XES, providing detailed information about the electronic, magnetic, orbital, and lattice excitations of materials. In a RIXS experiment, a monochromatic incident beam is scanned across an absorption edge while the energy of the scattered photons is measured with high spectral resolution. The resulting two-dimensional spectrum maps the energy transferred to the material as a function of incident photon energy, providing direct access to collective excitations such as magnons, phonons, plasmons, orbitons, and other low-energy quasiparticles in quantum materials including cuprates, nickelates, and iridates~\cite{kotani2001resonant, ament2011resonant, de2024resonant}.

An advantage of RIXS is its potential for high energy resolution. In contrast to XAS and XES, whose resolution is intrinsically limited by the finite lifetime of the core-hole states involved, RIXS is described by the Kramers--Heisenberg formalism, in which the final state generally does not contain a deep core hole. As a result, many RIXS features can have intrinsically narrow linewidths, and the achievable energy and momentum resolutions are often determined primarily by the instrumental resolution rather than by fundamental lifetime broadening. Consequently, the bandwidth of the incident X-ray beam plays an important role in determining the overall performance of a RIXS experiment. The narrow linewidth of an XFELO would therefore directly translate into improved spectroscopic resolution.

At the same time, RIXS is an intrinsically photon-hungry technique. Because radiative relaxation channels represent only a small fraction of all core-hole decay processes, the RIXS cross section is typically very small, making high-brightness X-ray sources essential. The higher spectral brightness of an XFELO, combined with its narrow bandwidth, would provide a dual advantage: increasing the usable photon flux while simultaneously improving energy resolution. These capabilities would reduce acquisition times, improve signal-to-noise ratios, and extend high-resolution RIXS measurements to previously difficult-to-study systems, thereby supporting studies of correlated quantum materials, catalytic processes, and complex chemical and biological systems.

\subsection{X-ray nonlinear spectroscopy}
\begin{figure*}[!hbt]
\centering
\includegraphics[width=0.9\linewidth]{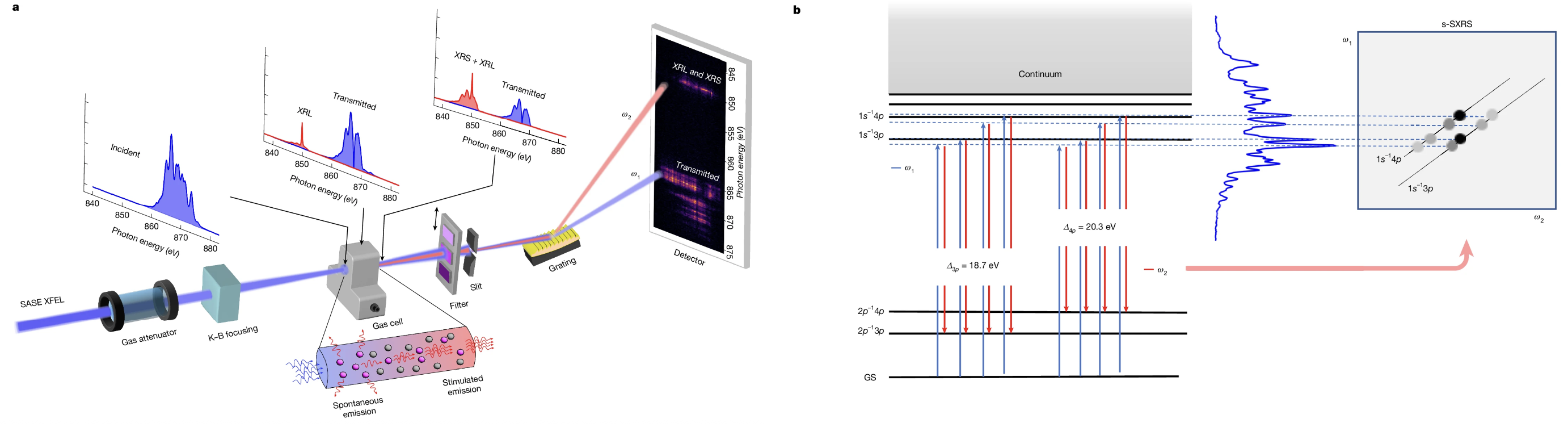}
\caption{\label{fig:sxrs} 
Super-resolution stimulated X-ray Raman spectroscopy~\cite{li2025super}.
}
\end{figure*}
Nonlinear spectroscopy employs multiple coherent light--matter interactions to probe couplings, coherence, and ultrafast dynamics that are inaccessible to conventional linear spectroscopy. It has become a widely used tool in modern physics, chemistry, and materials science, enabling applications ranging from label-free sub-micron imaging and the investigation of ultrafast carrier dynamics to the study of energy-transfer pathways, molecular vibrations, and the structure of molecular monolayers and biological membranes at liquid--solid and liquid--gas interfaces. Extending nonlinear spectroscopy into the X-ray regime would offer advantages. The short X-ray wavelength provides atomic-scale spatial resolution, while core-level transitions confer intrinsic element specificity. However, the limited temporal coherence and broad bandwidth of SASE XFELs restrict the spectral resolution and efficiency of many nonlinear X-ray spectroscopic techniques. The advent of XFELOs, with their narrow linewidth and high temporal coherence, could alleviate these limitations and improve the capabilities of nonlinear X-ray spectroscopy.

One of the fundamental building blocks of nonlinear X-ray spectroscopy is stimulated Raman spectroscopy (SRS). In the optical regime, SRS has been widely used to probe molecular vibrations, vibronic couplings, and energy-transfer dynamics. Extending SRS to the X-ray regime enables access to electronic excitations and ultrafast charge-transfer processes with elemental selectivity and atomic-scale resolution. In conventional X-ray Raman experiments, achieving high spectral resolution requires sophisticated monochromators for the incident beam and large, high-resolution spectrometers for the scattered photons. Although stimulated X-ray Raman scattering was demonstrated more than a decade ago, its spectroscopic potential remained largely unrealized because of the limited spectral resolution achievable with broadband SASE XFEL pulses~\cite{weninger2013stimulated,li2025super,linker2025attosecond}. Recently, as shown in Fig.~\ref{fig:sxrs}, covariance analysis combined with super-resolution techniques has significantly improved the energy resolution of stimulated X-ray Raman spectroscopy, revealing fine electronic states that were previously indistinguishable~\cite{li2025super,linker2025attosecond}. The narrow bandwidth and long coherence time of XFELO pulses would further enhance these capabilities by enabling high-resolution pump--probe measurements. Such advances would allow real-time tracking of ultrafast charge transfer and electronic dynamics in complex molecular systems with unprecedented spectral, temporal, and spatial resolution.

Beyond improving spectroscopic resolution, XFELOs could substantially enhance the signal strength of coherent nonlinear X-ray processes. Many nonlinear phenomena, including coherent multiphoton scattering such as harmonic generation and wave mixing, rely on phase coherence and efficient phase matching between interacting electromagnetic fields. The limited coherence time of SASE XFEL pulses, typically determined by individual temporal spikes with durations of only a few hundred attoseconds, restricts the coherent interaction time and reduces the efficiency of these nonlinear processes. For example, a recent demonstration of X-ray four-wave mixing (FWM) in neon gas employed a SASE XFEL source~\cite{morillo2026coherent}. The XFEL coherence time was significantly shorter than the core-hole lifetime, which is typically a few femtoseconds. As a result, the effective resonance linewidth was broadened by the limited coherence time of the incident radiation, reducing the nonlinear scattering cross section and the overall signal strength. An XFELO would largely eliminate this limitation, as its linewidth can be substantially narrower than the intrinsic core-hole lifetime broadening, allowing a larger fraction of the incident photons to participate in FWM and thereby increasing the signal by more than an order of magnitude. Furthermore, the long temporal coherence of XFELO radiation would enable efficient coherent interactions involving multiple pulses at different wavelengths, paving the way for advanced multicolor and multidimensional X-ray spectroscopies. These capabilities would extend many of the powerful techniques developed in optical nonlinear spectroscopy into the X-ray domain, providing unprecedented insight into electronic correlations, charge migration, and ultrafast quantum dynamics in complex systems.

\subsection{X-ray quantum optics}
One of the most remarkable phenomena in quantum optics is photon entanglement, which has played a central role in demonstrating the violation of Bell's inequalities and establishing the foundations of quantum information. Beyond fundamental physics, entangled photon pairs have found numerous applications in imaging and spectroscopy, enabling capabilities that are inaccessible with classical light sources. The strong correlations between entangled photons can reduce background, improve signal-to-noise ratios, and lower the radiation dose delivered to a sample. Furthermore, entangled photons enable entirely new spectroscopic approaches for probing light--matter interactions, particularly in complex biological and chemical systems~\cite{mukamel2020roadmap,sofer2019quantum,schlawin2018entangled}. Extending these concepts from the optical to the X-ray regime is especially attractive because X-rays provide intrinsic elemental specificity, atomic-scale spatial resolution, and access to core-electron excitations. As a result, X-ray entangled photons could open new opportunities for quantum-enhanced spectroscopy, imaging, and precision measurements of matter at the atomic scale.

\begin{figure}[!hbt]
\centering
\includegraphics[width=0.8\linewidth]{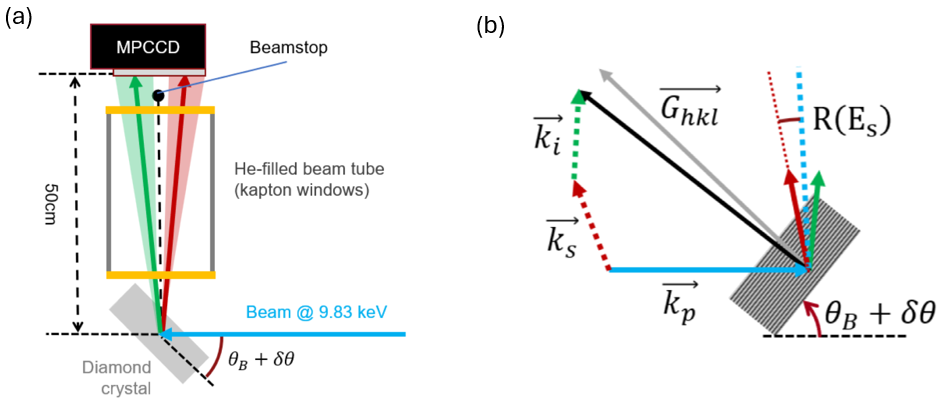}
\caption{\label{fig:spdc} 
(a) Schematic of the X-ray parametric down-conversion (XPDC) experimental setup.
(b) Energy and momentum conservation of XPDC in a crystal. Reproduced from~\citep{hartley2025x} under the \href{https://creativecommons.org/licenses/by/4.0/}{Creative Commons Attribution 4.0 International licence}.
}
\end{figure}

In the optical regime, entangled photon pairs are routinely generated through parametric down-conversion (PDC) in nonlinear optical crystals, where a pump photon is converted into two lower-energy photons, commonly referred to as the signal and idler photons. The process is governed by the conservation of both energy and momentum. Extending this mechanism to X-ray energies gives rise to X-ray parametric down-conversion (XPDC), in which nonlinear interactions within a crystal generate correlated X-ray photon pairs. As shown in Fig.~\ref{fig:spdc}, in contrast to optical PDC, phase matching in XPDC is achieved through momentum transfer to the crystal lattice, typically by operating slightly away from the exact Bragg condition to compensate for the dispersion of the interacting waves. Recently, entangled X-ray photon pairs were experimentally measured at the SACLA XFEL facility~\cite{hartley2025x}. The reported generation rate was approximately 1,250 photon pairs per hour, highlighting both the feasibility of the process with an XFEL and the challenge posed by its extremely low conversion efficiency. The substantially higher spectral brightness, repetition rate, and temporal coherence anticipated from XFELO sources could increase XPDC generation rates by several orders of magnitude, moving X-ray entangled-photon sources from proof-of-principle demonstrations toward practical tools for quantum-enhanced X-ray spectroscopy and imaging.

Quantum coherent control of nuclear transitions has long been considered impractical because of the extremely weak coupling between nuclear states and electromagnetic radiation, as well as the limited coherence and intensity of available X-ray light sources. The emergence of highly coherent X-ray lasers, particularly XFELOs, has changed this perspective. In recent years, several experiments have demonstrated coherent nuclear light--matter interactions. For example, Rabi oscillations of an X-ray photon between two resonant $^{57}$Fe layers embedded in two coupled cavities were observed~\cite{haber2017rabi}. These studies highlight the potential of coherent nuclear quantum optics in the collective regime.

Extending coherent control to individual nuclei remains a significant challenge. The stochastic phase and intensity fluctuations inherent to SASE XFEL pulses prevent the establishment of a well-defined coherent excitation and therefore strongly suppress phenomena such as nuclear Rabi oscillations and quantum coherent population manipulation. In contrast, the narrow linewidth and long temporal coherence of XFELO radiation could enable deterministic coherent interactions with nuclear transitions. Femtosecond X-ray pulses with near-transform-limited coherence and nanometer-scale focusing can achieve peak intensities approaching $10^{22}~\mathrm{W/cm^2}$~\cite{yamada2024extreme}, potentially sufficient to drive Rabi oscillations within the pulse duration for suitable nuclear transitions~\cite{burvenich2006nuclear}, opening the possibility of direct quantum control of nuclear states.

Beyond serving as a proof of principle, coherent manipulation of nuclear transitions would provide methods for precision measurements of nuclear properties, including transition energies, dipole moments, and coherence and excitation lifetimes. More broadly, it would establish the foundation for the emerging field of nuclear quantum optics, in which concepts traditionally associated with atomic and optical physics, such as coherent control, quantum interference, cavity quantum electrodynamics, and quantum information processing, and are extended to the nuclear domain. Such developments could support frequency standards, quantum sensors, and precision tests of fundamental physics.

\subsection{Precision X-ray metrology}
Modern precision metrology and laser spectroscopy are predominantly based on electronic transitions in atoms and ions. However, long-lived nuclear isomeric states, with resonance quality factors comparable to or even exceeding those employed in current optical atomic clocks, offer advantages due to their resilience against external perturbations. These properties make nuclear transitions promising candidates for future frequency standards and could advance quantum metrology, clock technology, chronometric geodesy, and gravimetry. Furthermore, nuclear clocks could enable sensitive tests of fundamental physics that rely on high-precision frequency measurements, including searches for temporal and spatial variations of fundamental constants, dark matter interactions, violations of Lorentz invariance, and tests of Einstein's equivalence principle~\cite{hayes2007sensitivity,peik2021nuclear,fadeev2020sensitivity,fuchs2025searching,arakawa2026probing,Fortier:26}.

Realizing these applications requires the direct optical excitation of nuclear transitions using high-brightness coherent light sources. Most nuclear transitions lie in the hundreds-of-keV to MeV energy range and are therefore inaccessible to conventional laser systems. A remarkable exception is the low-energy isomeric transition in $^{229}$Th, whose energy of approximately 8.4 eV (148 nm) allows direct excitation with vacuum-ultraviolet (VUV) lasers. This transition has recently been studied extensively, and a nuclear clock based on it has been demonstrated~\cite{tiedau2024laser,elwell2024laser,zhang2024frequency,morawetz2026continuous,de2026thorium,huang2026a}. Beyond $^{229}$Th, numerous nuclear transitions reside in the X-ray regime, where XFELs provide a unique platform for resonant excitation and detection. One particularly attractive example is the long-lived nuclear transition in $^{45}$Sc, which possesses an exceptionally high quality factor. Recently, resonant excitation of the $^{45}$Sc nuclear transition was successfully demonstrated at the European XFEL~\cite{shvyd2023resonant}. By detecting the fluorescence following excitation, the transition energy was determined to be 12,389.59 eV, reducing the uncertainty by approximately two orders of magnitude compared with previous measurements. Despite this achievement, the excitation probability remains extremely low, limiting the experimental efficiency, while the attainable energy resolution is constrained by the bandwidth of the incident X-ray pulses. The advent of high-brightness, narrow-linewidth XFELOs could significantly enhance both the excitation efficiency and spectroscopic precision of such experiments, opening new opportunities for nuclear quantum optics and precision nuclear spectroscopy.

\begin{figure}[!hbt]
\centering
\includegraphics[width=0.8\linewidth]{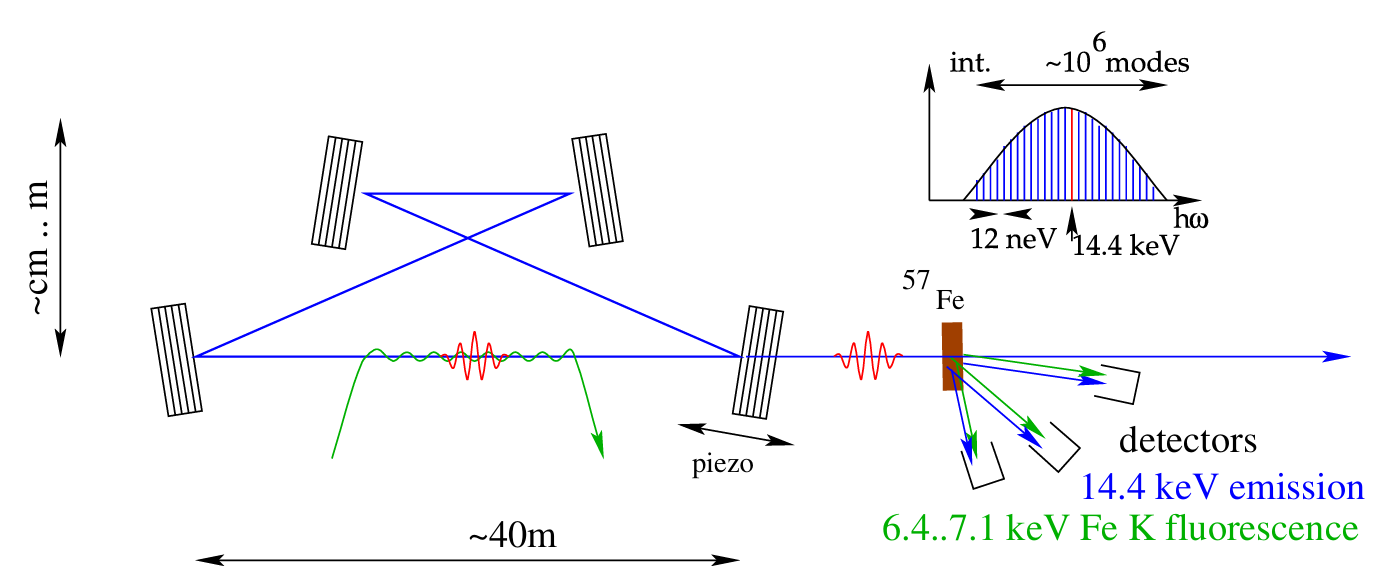}
\caption{\label{fig:comb} 
Schematic of X-ray comb generation based on an XFELO. Reproduced from~\cite{adams2015x} under the \href{https://creativecommons.org/licenses/by/3.0/}{Creative Commons Attribution 3.0 Unported licence}.
}
\end{figure}

Since the invention of the laser, sustained efforts have been devoted to improving its stability, coherence, and spectral purity. An important development was the optical frequency comb, which established a phase-coherent link between optical and microwave frequencies and enabled advances in optical atomic clocks and precision laser spectroscopy, an achievement recognized by the 2005 Nobel Prize in Physics. The advent of the XFELO offers the possibility of extending these techniques into the X-ray regime. Unlike conventional SASE XFELs, an XFELO employs an optical feedback cavity formed by high-reflectivity Bragg crystals, providing a stable seed for successive electron bunches and improving the temporal coherence, spectral stability, and brightness of the emitted radiation. As shown in Fig.~\ref{fig:comb}, by stabilizing the XFELO cavity length using laser-based locking techniques and potentially ultranarrow nuclear transitions as frequency references, a pulsed XFELO could generate an X-ray frequency comb with phase coherence maintained between successive pulses~\cite{adams2015x}. Such a source could improve the coherence and reduce the spectral linewidth by more than six orders of magnitude compared with existing XFEL facilities, reaching the neV regime.

The combination of high brightness, ultranarrow linewidth, and long-term frequency stability would advance high-resolution X-ray spectroscopy and precision metrology. In particular, X-ray frequency combs could enable high-resolution nuclear M\"ossbauer spectroscopy, improving the precision with which nuclear transition energies, hyperfine interactions, and nuclear structure parameters are measured, thereby providing quantitative benchmarks for nuclear theory. More broadly, highly coherent X-ray sources could extend many of the established techniques developed for optical precision measurements, including interferometry, frequency metrology, and precision sensing, into the X-ray domain. In the long term, such capabilities may support fundamental physics experiments and applications in gravitational-wave detection, dark-matter searches, and searches for temporal variations in fundamental constants, analogous to the impact that optical frequency combs and laser interferometry have had on modern precision science.

\section{Challenges and Future Directions}
\label{sec:future}

CBXFELs now have an experimentally credible basis as coherent, high-repetition-rate, narrow-bandwidth X-ray sources \citep{Kim2008,HuangRuth2006,Adams2019,Rauer2026}. Multi-pass gain in a Bragg cavity has been demonstrated, but reproducible user operation of the coupled accelerator, undulator, and optical cavity system has not. The features that make XFELOs and RAFELs attractive also impose their principal constraints. High-finesse Bragg optics have narrow angular and spectral acceptance. Repeated electron--photon interactions accumulate timing, energy, and alignment errors, while high average flux converts weak absorption into sustained thermal load. A practical CBXFEL must preserve the circulating mode, control heat deposition, maintain phase coherence, and extract usable power. It must also diagnose the stored field without disrupting the feedback that sustains oscillation.

\subsection{Alignment, diagnostics, and feedback}

The European XFEL experiment demonstrated synchronization of a long hard-X-ray Bragg cavity with a superconducting-linac bunch train and measured multi-pass gain \citep{Rauer2026}. It also clarified the remaining control problem. A micrometer-scale X-ray mode and each fresh electron bunch must overlap throughout a multi-element cavity extending tens to hundreds of meters. Angular drift, mirror figure errors, thermal strain, and beam jitter are coupled. Each perturbation can alter gain, round-trip loss, wavefront quality, and spectral detuning \citep{QiShvydko2022,Margraf2023LowLoss,Liu2024Optics}.

Nondestructive diagnostics are therefore integral to the oscillator rather than auxiliary beamline equipment. The stored field must be sampled sensitively while adding negligible loss. Required observables include transmission, wavefront, spectrum, pointing, timing, and pulse-train evolution. Recent diagnostic studies target sub-microradian angular alignment, micrometer-scale spatial overlap, low-loss sampling, and single-shot spectral readout with meV resolution \citep{Liu2024Diagnostics,Kauchha2025Spectrograph}. These capabilities are needed for both commissioning and routine operation. User facilities would additionally require automatic drift recovery, machine-protection thresholds, and reproducible switching among spectral, temporal, and out-coupling modes.

The feedback architecture remains an open systems problem because its control variables evolve on different time scales. These variables include electron-beam energy, arrival time, undulator settings, cavity length, crystal angle, thermal state, and out-coupling. A plausible hierarchy would combine fast timing and energy feedback with slower Bragg-angle and cavity-length loops. Supervisory control would then limit mode hopping and thermal runaway. The performance of this architecture remains to be demonstrated in sustained operation.

\subsection{High average power and thermal management}

High average power is both a primary motivation and a major materials challenge for CBXFELs. Superconducting linacs and energy-recovery linacs can supply high-repetition-rate electron beams, while a cavity can reuse a coherent seed over many passes \citep{Decking2020,Galayda2018LCLSII,Wang2020Energy}. This architecture is designed to exceed the average spectral brightness of single-pass SASE sources. It also shifts the expected limitation from single-pulse gain toward accumulated thermal deformation, mechanical drift, and radiation damage in X-ray optics \citep{HuangDeng2020Thermal,Liu2024Thermoelastic}.

The relevant figure of merit is absorbed rather than delivered power. In a low-gain XFELO, high cavity $Q$ and weak extraction can make intracavity power much larger than user power. A RAFEL permits stronger extraction, although its amplified pulse can still deposit substantial heat. At MHz repetition rates, even weak absorption can alter lattice spacing and shift the Bragg condition. It can also broaden the reflectivity curve, degrade transverse-mode quality, and introduce phase-front errors \citep{HuangDeng2020Thermal,Liu2024Thermoelastic,Shvydko2021Diamond}. Thermal management is therefore as much a spectral-stability problem as a heat-removal problem.

High-gain operation changes this optimization. Low-gain XFELOs require very low round-trip loss and weak output coupling because their single-pass gain is modest. RAFELs can operate with a smaller feedback fraction because one high-gain pass replenishes the seed \citep{HuangRuth2006,Freund2019,Marcus2020HighPower}. This approach relaxes the cavity-$Q$ requirement but demands tighter regulation of pulse energy, heat deposition, gain saturation, and wavefront quality. Marcus et al. proposed a high-power RAFEL with undulator tapering and an apertured diamond out-coupler \citep{Marcus2020HighPower}. In their design, post-saturation refractive guiding limits the stored cavity power. The broader design principle is to retain enough radiation for seeding without thermally detuning the Bragg optics.

Future high-average-power sources will require integrated thermal control. Low-absorption diamond optics, effective heat sinking, and cryogenic or actively temperature-stabilized operation must work with adaptive cavity and beam-based feedback. Recent component studies have characterized high-reflectivity diamond crystals, drumhead membranes, refractive lenses, wavefront-preserving diagnostics, and meV-resolution spectrographs \citep{Liu2024Optics,Liu2024Diagnostics,Kauchha2025Spectrograph}. End-to-end models should couple FEL gain and Bragg diffraction to heat flow, acoustic vibration, mirror errors, radiation damage, and long-term drift \citep{HuangDeng2020Thermal,Liu2024Thermoelastic,Krzywinski2022BPM}. The target is a self-consistent steady state throughout long pulse trains. At that point, gain saturation, out-coupling, Bragg response, and thermal deformation must remain mutually compatible.

\subsection{Pulse shaping and structured X-ray fields}

A major scientific opportunity is to extend CBXFELs from narrow-band coherent oscillators to programmable X-ray sources. The difficulty is intrinsic to the cavity. Bragg reflection provides spectral filtering and mode cleaning, but its narrow acceptance suppresses broadband, few-femtosecond, and arbitrarily shaped waveforms \citep{Shvydko2012Spatiotemporal,Lindberg2012Time,Zhang2025PulseShaping}. Any pulse-shaping scheme must therefore add structure without removing the radiation from the high-reflectivity cavity mode.

Transverse structuring offers one proposed route because the cavity eigenmode can, in principle, be engineered. Huang and Deng proposed using Bragg mirrors and longitudinal--transverse mode coupling to generate orbital-angular-momentum (OAM) X-rays in an XFELO \citep{Huang2020OAM}. Related eigenmode designs could support vortex beams, higher-order transverse modes, switchable OAM states, and tailored polarization \citep{Huang2019Polarization,Huang2023Rapidly,Yan2022Self}. Potential applications include chiral spectroscopy, magnetic scattering, angular-momentum transfer, and studies of topological materials. Modal selectivity remains the central limitation. Hard-X-ray Bragg reflection couples angle, frequency, wavefront, and lattice strain. Structured-field operation must therefore suppress mode competition while preserving reflectivity, round-trip gain, and wavefront quality.

Temporal shaping faces a stronger constraint from the Bragg bandwidth. A meV-scale hard-X-ray cavity naturally selects subpicosecond to picosecond pulses, whereas many ultrafast applications require femtosecond or attosecond structure \citep{Shvydko2012Spatiotemporal,Lindberg2012Time}. Proposed RAFEL architectures separate the narrow-band stored seed from reshaping in the high-gain pass. In a laser-modulation-driven RAFEL, a shaped optical laser imprints an energy modulation on the electron beam. The modulated bunch generates sidebands outside the Bragg bandwidth, allowing them to exit while the original seed remains trapped \citep{Zhang2025PulseShaping}. This proposed mechanism combines pulse shaping, color control, and output coupling in one process.

The modeled control variable is slice-dependent FEL detuning. Energy or current modulation determines which longitudinal slices remain resonant, which sidebands grow, and how gain saturation redistributes power. High single-pass gain can then amplify selected temporal or spectral components, while the cavity retains longitudinal coherence and memory \citep{Zhang2025PulseShaping,Li2017High}. The next step is an experimental demonstration of shaped output from a cavity-based system. It must show that programmability coexists with stable build-up, manageable heat load, and shot-to-shot reproducibility.

\subsection{Mode-locked X-ray FELs}

\begin{figure}[hbt!]
    \centering
    \includegraphics[width=\linewidth]{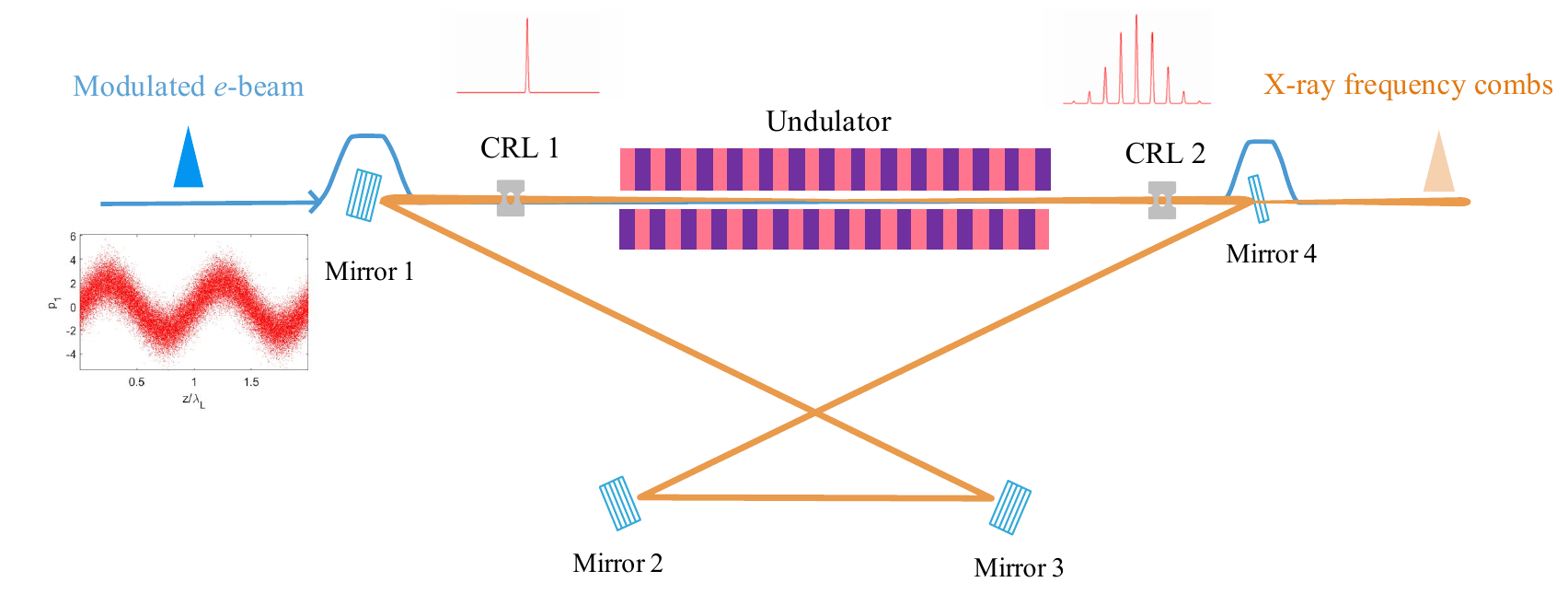}
    \caption{Concept of an actively mode-locked CBXFEL. In the proposed scheme, a modulated electron beam interacts with the intracavity monochromatic X-ray field in the main undulator, producing a phase-locked pulse train (frequency comb). Reproduced from~\citep{Huang2025ModeLock} under the \href{https://creativecommons.org/licenses/by/4.0/}{Creative Commons Attribution 4.0 International licence}.}
    \label{fig:ML_XFELO}
\end{figure}

Mode locking could extend CBXFELs from narrow-band oscillators to hard-X-ray pulse-train and comb sources. The objective is to phase-lock multiple longitudinal modes, producing short temporal pulse trains and comb-like spectra. Proposed applications include attosecond pump--probe measurements, high-resolution spectroscopy, X-ray quantum optics, and precision metrology \citep{Huang2025ModeLock,Hu2025ModeLockedComb,Liu2026Phase}. Unlike in optical lasers, modulation could be applied through the electron beam rather than an intracavity X-ray modulator. Figure~\ref{fig:ML_XFELO} sketches this proposed actively mode-locked CBXFEL.

Recent proposals and a related single-pass experiment outline a possible path. Huang, Yang, and Deng proposed coherent electron-beam energy modulation for an actively mode-locked CBXFEL \citep{Huang2025ModeLock}. Their three-dimensional simulations predict $700~\mu{\rm J}$ total energy, $30~{\rm GW}$ peak power, and $1.55~{\rm eV}$ comb spacing set by the modulation frequency. The simulated tolerance to large reflectivity variations suggests that active modulation may ease some optical-quality constraints. This tolerance comes at the cost of stricter timing and synchronization requirements. Zhang et al. independently proposed a similar scheme~\cite{Zhang2025PulseShaping}.

A single-pass X-ray FEL frequency-comb experiment provides complementary evidence. Magnetic chicanes and an external optical laser confined gain to periodic regions of the electron bunch. The experiment produced phase-locked subfemtosecond pulse trains and a comb-like spectrum \citep{Hu2025ModeLockedComb}. Although it was not a cavity oscillator, it demonstrated phase-correlated X-ray emission through electron-beam modulation and mode coupling. It therefore validates the central single-pass mechanism, but not phase preservation over repeated cavity round trips. Future CBXFEL mode locking should be verified jointly in spectral, temporal, and electron-beam observables rather than inferred from one signature.

The unresolved issue is phase preservation over many round trips. Timing jitter shifts the modulation phase, energy jitter alters detuning, and cavity-length jitter displaces the longitudinal-mode comb. Their accumulation degrades comb coherence. Proposed mode-locked CBXFELs therefore require subfemtosecond timing diagnostics, optical-to-RF synchronization, and feedback on cavity length and electron-beam energy \citep{Stoupin2010Nanoradian,Koehlenbeck2024Dynamic,Liu2026Phase}. They must also meet a bandwidth compromise. The spectrum must be broad enough to form short pulses but narrow enough for the Bragg cavity to preserve phase relationships.

\subsection{Out-coupling and cavity tuning}

\begin{figure}[hbt!]
    \centering
    \includegraphics[width=\linewidth]{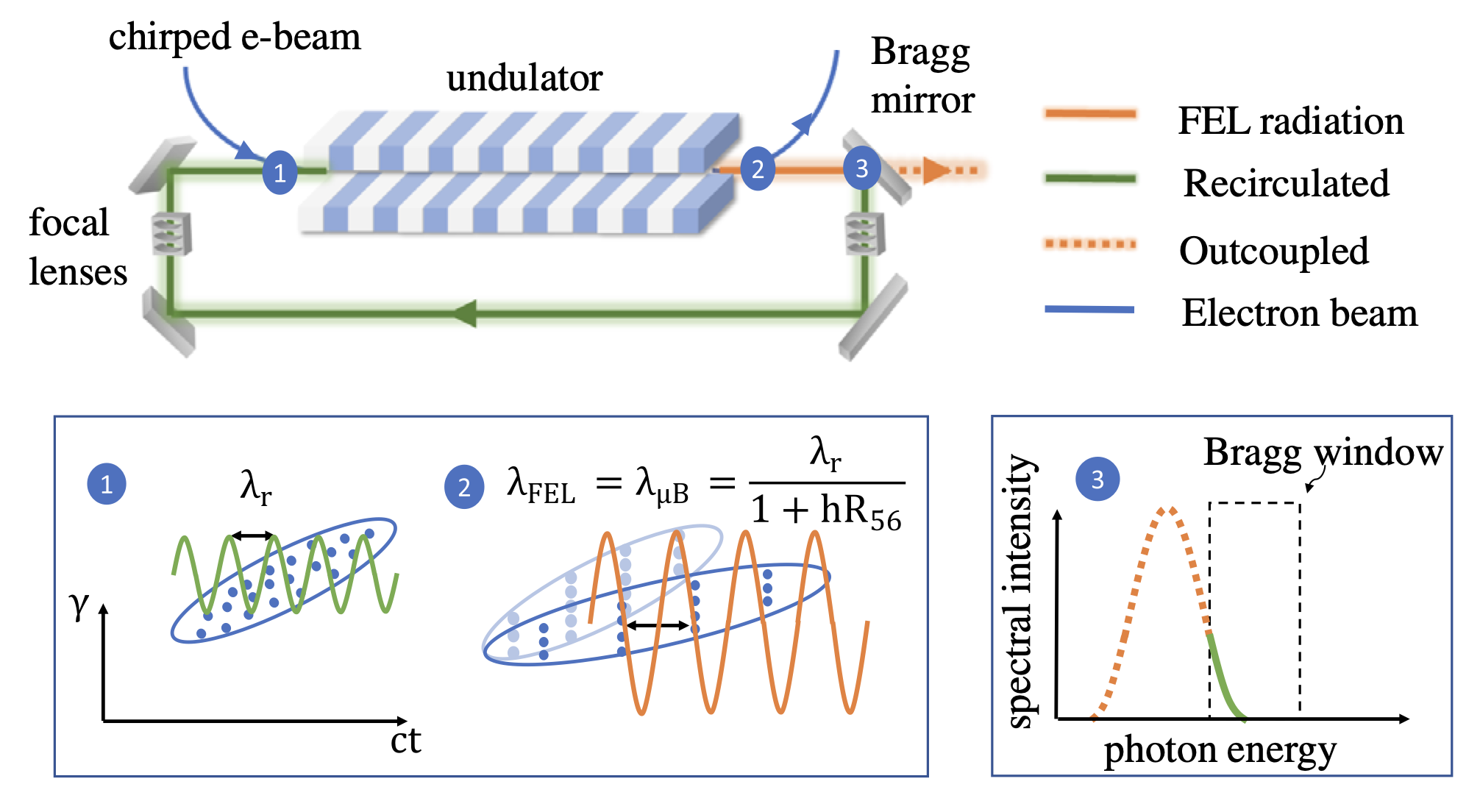}
    \caption{Schematic of proposed beam-controlled Q-switching in an X-ray RAFEL. An electron-bunch energy chirp modifies longitudinal compression and tunes the microbunching and radiation wavelengths, $\lambda_{\mu B}$ and $\lambda_{\mathrm{FEL}}$. The shift places most of the FEL spectrum outside the Bragg-reflection band for extraction. The matched fraction remains in the cavity to seed the next pass. Reproduced from~\citep{Tang2023} under the \href{https://creativecommons.org/licenses/by/4.0/}{Creative Commons Attribution 4.0 International licence}.} 
    \label{fig:qswitching}
\end{figure}

Out-coupling links oscillator physics, user flux, and thermal management. Optical lasers can use partially transmitting mirrors, whereas hard-X-ray cavities rely on Bragg crystals with narrow spectral and angular acceptance. A thin Bragg crystal can act as a permeable mirror, but extraction directly consumes the loss budget required for oscillation. This constraint is severe for low-gain XFELOs. High-gain RAFELs can support larger extraction fractions \citep{Shvydko2019Outcoupling,Freund2019,Marcus2020HighPower,Tiwari2025PowerLoss}. Figure~\ref{fig:qswitching} shows one proposed beam-controlled Q-switching route to active extraction.

Three approaches define the current design space. First, intracavity diamond beam splitters can combine Bragg reflection with X-ray transmission, providing adjustable and multibeam out-coupling \citep{Shvydko2019Outcoupling}. Diffraction, apertures, lenses, and Bragg-reflection losses limit their useful range \citep{Tiwari2025PowerLoss}. Second, Q-switching treats the effective cavity quality factor as a beam-controlled variable. Tang et al. proposed regulating RAFEL output through electron-beam phase-space manipulation \citep{Tang2023}. Third, proposed sideband extraction keeps the seed within the Bragg bandwidth while transmitting modulation-generated sidebands \citep{Zhang2025PulseShaping}. This approach avoids additional intracavity X-ray optics but couples waveform design to detuning, gain saturation, and modulation stability.


\subsection{CBXFELs at fourth-generation synchrotrons}

\begin{figure}[hbt!]
    \centering
    \includegraphics[width=\linewidth]{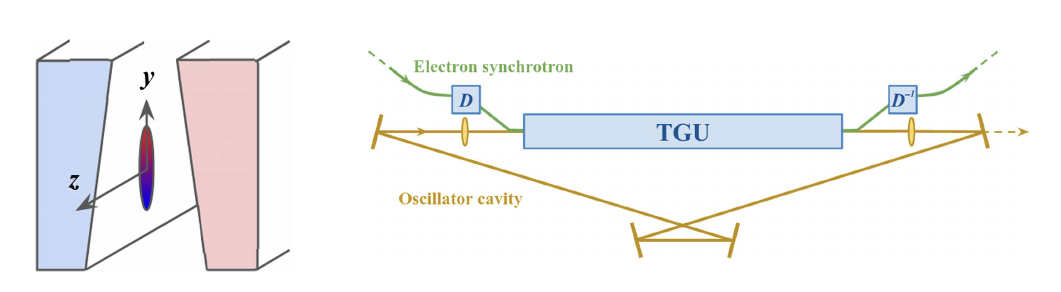}
    \caption{Left: transverse-gradient undulator. Right: proposed layout of a synchrotron-driven TGU-XFELO. The system comprises an electron synchrotron, a transverse-gradient-undulator section, and an X-ray optical cavity. Two dispersive segments, $D$ and $D^{-1}$, provide dispersion matching and recovery around the TGU. Reproduced from~\citep{Li2023TGU} under the \href{https://creativecommons.org/licenses/by/4.0/}{Creative Commons Attribution 4.0 International licence}.}
    \label{fig:TGU_XFELO}
\end{figure}

Fourth-generation synchrotrons, or diffraction-limited storage rings, are candidate CBXFEL hosts. Their multibend-achromat lattices combine low equilibrium emittance, high average current, orbit stability, and mature multi-user infrastructure \citep{Tavares2014MAXIV,Shin2021NewEra,Raimondi2023EBS,Schroer2019PETRAIV}. Their limitation is longitudinal coherence because spontaneous storage-ring undulator radiation is not a regenerated laser field. A proposed ring-based XFELO would use the storage ring as a high-repetition-rate electron source and a Bragg cavity as spectral memory \citep{Kim2008,Lindberg2011XFELOPerformance,Li2023TGU}. The central constraint is low-gain phase-space density rather than average current. Single-pass gain must exceed round-trip cavity loss within the narrow Bragg bandwidth. Meanwhile, the equilibrium energy spread and bunch length remain large relative to the scales required for hard-X-ray FEL gain \citep{KimShvydko2009,Lindberg2011XFELOPerformance,Lindberg2013TGUStorageRing}. Figure~\ref{fig:TGU_XFELO} shows a representative proposed TGU-XFELO layout.

The transverse-gradient-undulator (TGU) approach directly targets this gain dilution. The original concept combined dispersion with a transverse magnetic-field gradient to reduce FEL sensitivity to electron-energy spread \citep{Smith1979TGU,Kroll1981TGU}. Later studies extended it to short-wavelength high-gain FELs and diffraction-limited-ring designs \citep{Huang2012CompactTGU,Baxevanis2014TGU3D,Lindberg2013TGUStorageRing}. A recent three-dimensional low-gain analysis of a storage-ring TGU-XFELO \citep{Li2023TGU} indicates that useful gain is possible for fourth-generation-ring parameters only within a tight multiparameter compromise. The relevant variables include dispersion, beam size, optical-mode overlap, emittance ratio, and ring-FEL coupling. Harmonic lasing and strong focusing add flexibility for 3-4~GeV rings by shifting hard-X-ray operation to higher undulator harmonics. However, they place harmonic selection, transverse matching, energy spread, bunch length, and cavity loss within the same narrow optimization window \citep{Dai2011,Yu2024SmallGain,Yu2023Feasibility}.

Storage-ring CBXFELs would complement rather than replace superconducting-linac systems. Linacs provide fresh, high-brightness bunches with low slice energy spread. These properties favor high-gain demonstrations, RAFEL operation, and high pulse energy. Storage rings offer continuous high-repetition-rate operation and facility stability. However, a dedicated FEL insertion must preserve beam lifetime and routine user operation while adding local dispersion, tight focusing, a long resonator, and feedback. Earlier storage-ring FEL studies showed macrotemporal oscillations and chaotic behavior when damping, gain, and cavity loss coupled unfavorably \citep{Elleaume1984Macrotemporal,Billardon1990Chaos}. Short-wavelength analyses also identify energy-spread growth, intrabeam scattering, coherent synchrotron radiation, dynamic aperture, and equilibrium beam degradation as design constraints \citep{Huang2008StorageRingFEL,Cai2025RingFELDynamics}.

The optical cavity completes the same systems problem. A ring-based XFELO requires low round-trip loss, stable focusing, controlled output coupling, and thermal robustness at modest single-pass gain \citep{KimShvydko2009,Lindberg2011XFELOPerformance}. Recent studies indicate that optimized diamond-mirror resonators may tolerate percent-level losses while providing weak but useful output coupling \citep{Li2023TGU,Tiwari2025PowerLoss}. This conclusion depends on co-designing crystal quality, focusing optics, apertures, heat load, alignment, and the electron lattice. If these conditions can be met, the likely niche is not femtosecond peak power. It is stable, high-average-rate, narrow-band, longitudinally coherent X-rays in a synchrotron-like user environment. Candidate applications include high-resolution spectroscopy, nuclear resonant scattering, correlation spectroscopy, and weak nonlinear X-ray measurements \citep{Adams2019,Shin2021NewEra,Raimondi2023EBS}.

\section{Conclusions and Outlook}
\label{sec:conclusion}

Realizing a user-ready CBXFEL essentially requires solving a tightly coupled three-body problem involving the relativistic electron beam, X-ray gain dynamics, and the high-finesse Bragg crystal cavity. To date, we have resolved the three constituent two-body interactions: the electron beam and X-ray amplification underpin all conventional XFEL operations and are now quantitatively understood from first-principles theory to saturated high-gain experiments; X-ray gain and the X-ray cavity have been validated via low-loss storage of hard X-rays in meter-scale Bragg resonators, showing that cavity optics can sustain high-brightness radiation without substantial degradation; and the electron beam and X-ray cavity have been synchronized to micrometer-scale spatial and femtosecond-scale temporal precision via active stabilization and passive retro-reflection geometries. The recent lasing demonstration at the European XFEL marks the first observation of the full three-body interaction: a 132.8~m Bragg cavity recirculated a 6.952~keV pulse and synchronized it to a 2.23~MHz bunch train \citep{Rauer2026}. Fresh electron bunches then amplified the pulse over successive passes. Together with low-loss storage tests, this result confirms that FEL gain, Bragg filtering, cavity transport, and bunch-to-bunch synchronization can operate as one system \citep{Kim2008,HuangRuth2006,Margraf2023LowLoss}. However, this evidence does not yet establish user-ready CBXFEL operation. The long-term goal for the field is to control this coupled system to achieve stable user operation.

The useful CBXFEL design space is best organized by gain and feedback. Low-gain XFELOs prioritize narrow-band storage and mode selection. High-gain RAFELs permit stronger extraction but are more sensitive to saturation, beam jitter, and nonlinear wavefront evolution. Oscillator-seeded amplifiers and pulse-shaping schemes occupy an intermediate regime, where the cavity preserves spectral memory while the undulator reshapes the delivered pulse. Across these architectures, controlled optical memory is the central unresolved problem. The recirculated field preserves coherence but also accumulates timing, energy, alignment, thermal, and output-coupling errors.

Absorbed power in Bragg optics is likely to set the dominant engineering limit at megahertz repetition rates \citep{HuangDeng2020Thermal,Liu2024Thermoelastic,Rauer2026}. CBXFELs must therefore be optimized for stable operating points rather than maximum small-signal gain or reflectivity. Progress requires low-defect crystals, thermal control, low-loss focusing, controlled out-coupling, pulse-resolved spectroscopy, nondestructive diagnostics, and coordinated accelerator--cavity feedback \citep{Liu2024Optics,Liu2024Diagnostics,Kauchha2025Spectrograph}. Future tests should progress from passive storage to multi-pass gain, deliberate extraction, long-pulse-train stability, and recovery from controlled perturbations. They must also quantify linewidth, longitudinal coherence, phase stability, transverse-mode purity, spectral reproducibility, and delivered flux.

An intrinsic trade-off in CBXFELs is that the high temporal coherence afforded by dynamical Bragg diffraction comes at the cost of restricted wavelength agility, producing the well-established ``deadband problem''.
Far from being an isolated optical artifact, this restriction is a direct manifestation of the tightly coupled three-body interaction. Near-normal incidence is mandatory for high reflectivity in hard X-rays, rendering Bragg resonance extremely sensitive to sub-$\mu$rad angular deviations that challenge current nanopositioning systems. In addition, lasing-induced thermal lattice strain and refractive index gradients lock the cavity to a fixed wavelength post-saturation, with post-tuning re-stabilization taking minutes to hours, orders of magnitude slower than millisecond-scale SASE tuning. Overcoming this limitation requires either alternative crystal-mounting or cryogenic schemes to decouple thermal expansion from lattice spacing, or hybrid architectures in which a narrow-band CBXFEL seeds a tunable tapered amplifier \citep{DengFeng2013,Marcus2020HighPower}. The latter aligns naturally with SHINE/MING's downstream long undulator sections reserved for high-efficiency power extraction.

It is worth noting that prior to the full maturation of hard X-ray systems, cavity-based FELs operating in the extreme-ultraviolet (EUV) and beyond-extreme-ultraviolet (BEUV) regimes may serve as critical technological precursors. Driven by the urgent demand for next-generation EUV/BEUV lithography, developments in Mo/Si multilayer mirror technology offer a viable path forward. Owing to the reduced system tolerances and mature coating techniques at EUV/BEUV wavelengths, cavity-based EUV/BEUV FELs, especially RAFELs, provide a pragmatic platform for demonstrating high-finesse cavity stabilization, active feedback control, and long-term operational reliability. Notably, an EUV-band RAFEL is among the top-priority inaugural experiments under active development by the SHINE collaboration, leveraging the flexible configuration of the facility's FEL-II undulator line. The expertise gained from EUV-scale implementations will accelerate the translation of cavity-based concepts to the challenging X-ray domain.

Meeting these requirements could fill a source gap between SASE XFELs, synchrotrons, and conventional self-seeding. Likely beneficiaries include high-resolution spectroscopy, nuclear resonant scattering, nonlinear X-ray methods, quantum optics, and precision metrology. Longer-term proposals add mode locking, frequency combs, structured eigenmodes, and shaped high-energy output \citep{Huang2025ModeLock,Hu2025ModeLockedComb,Zhang2025PulseShaping}. These advanced capabilities remain proposals rather than demonstrated features of an integrated cavity source. The decisive question is whether accelerator stability, X-ray optics, thermal control, diagnostics, and feedback can jointly deliver reproducible coherent X-ray modes. Until then, the broad scientific impact of CBXFELs remains a credible prospect rather than an established capability.

\section*{Acknowledgments}

The authors acknowledge the support and collaboration of colleagues in the field of XFEL physics, facilities, and science applications. This work was supported by the National Natural Science Foundation of China (12125508, 12541503), the National Key Research and Development Program of China (2024YFA1612104), and Shanghai Pilot Program for Basic Research -- Chinese Academy of Sciences, Shanghai Branch (JCYJ-SHFY-2021-010). Patrick Rauer acknowledges support from DESY (Hamburg, Germany), a member of the Helmholtz Association (HGF), and the European XFEL (Schenefeld, Germany).

\clearpage
\bibliographystyle{elsarticle-num}
\bibliography{refs}
\end{document}